\documentclass[jcp,aps,showpacs,twocolumn,supergroupedaddress,epsfig,amsmath,amssymb,eqsecnum]{revtex4}
\usepackage{epsfig,amsmath,amssymb,bm,epsf,graphics,psfrag,verbatim,subfigure,framed}
\usepackage[usenames,dvipsnames]{color}
\usepackage{bbm}
\usepackage{mathrsfs}
\usepackage{amsfonts}
\usepackage{color}
\usepackage{pbox}
\def\tA{\widetilde{A}}
\def\tB{\widetilde{B}}

\newcommand{\be}{\begin{equation}}
\newcommand{\ee}{\end{equation}}
\newcommand{\ba}{\begin{eqnarray}}
\newcommand{\ea}{\end{eqnarray}}
\newcommand{\bw}{\begin{widetext}}
\newcommand{\ew}{\end{widetext}}

\newcommand{\xv}{{\mathbf{x}}}

\newcommand{\bing}[1]{\textcolor{black}{#1}}

\begin{document}

\title{Van der Waals torque and force between dielectrically anisotropic layered media}
\author{Bing-Sui Lu$^1$}
\email{binghermes@gmail.com}
\author{Rudolf Podgornik$^{1,2}$}
\affiliation{$^{1}$Department of Theoretical Physics, J. Stefan Institute, 1000 Ljubljana, Slovenia}

\affiliation{$^{2}$Department of Physics, Faculty of Mathematics and Physics, University of Ljubljana, 1000 Ljubljana, Slovenia.}
\date{\today}

\begin{abstract}
We analyse \bing{van der Waals interactions between} a pair of dielectrically \emph{anisotropic} plane-layered media interacting across a dielectrically isotropic solvent medium. 
\bing{We develop a general formalism based on transfer matrices to investigate the van der Waals torque and force
in the limit of weak birefringence and dielectric matching between the ordinary axes of the anisotropic layers and the solvent.} \bing{We apply this formalism to study the following systems:} (i)~a pair of single anisotropic layers, (ii)~a single anisotropic layer interacting with a multilayered slab consisting of alternating anisotropic and isotropic layers, and (iii)~a pair of multilayered slabs each consisting of alternating anisotropic and isotropic layers, looking at the cases where the optic axes lie parallel and/or perpendicular to the plane of the layers. For the first case, the optic axes of the oppositely facing anisotropic layers of the two interacting slabs generally possess an angular mismatch, and within each multilayered slab the optic axes may either be the same, or undergo constant angular increments across the anisotropic layers. In particular, we examine how the behaviors of the van der Waals torque and force can be ``tuned" by adjusting the layer thicknesses, the relative angular increment within each slab, and the angular mismatch between the slabs. 
\end{abstract}

\maketitle

\section{Introduction}

Van der Waals (vdW) forces exist between any pair of bodies if their material polarizability differs from the background~\cite{parsegian-vdw,mahanty,sernelius,french,woods}. Additionally, a vdW \emph{torque} can appear if these bodies  \bing{display either an anisotropic shape or are} \emph{birefringent} \cite{hopkins}, i.e., their dielectric properties are different along different principal dielectric axes, as is typically the case with crystals such as quartz or crystallite structures such as kaolinite~\cite{born-wolf}. In dielectrically (or optically) anisotropic materials, there is a special principal axis called the {\em optic axis}, which coincides with the axis of symmetry of the \emph{dielectric ellipsoid} of the crystal (see Fig.~\ref{fig:anisotropy}).  Dielectrically anisotropic materials can be classified as either \emph{uniaxial} or \emph{biaxial}, depending on whether the principal dielectric permittivities in the directions perpendicular to the optic axis are respectively identical or distinct~\cite{born-wolf,landau}. 

The dielectric anisotropy effects were first addressed in the Lifshitz theory of vdW interactions for isotropic boundaries and anisotropic intervening material by Kats \cite{Kats,Kornilovitch}, while Parsegian and Weiss independently formulated the non-retarded Lifshitz limit for vdW torques in the case of two uniaxial half-spaces separated by another dielectrically anisotropic medium~\cite{parsegian-weiss}. Later the complete Lifshitz result, including retardation, for two dielectrically anisotropic half-spaces with an intervening isotropic slab was obtained by Barash~\cite{barash1,barash2,Capasso1, Capasso-erratum}. The general Lifshitz theory results for the vdW interactions in stratified anisotropic and optically active media with retardation effects are algebraically unwieldy \cite{Veble2}, not permitting any final simplification \cite{Philbin}. Further efforts in the investigation of the vdW torque between a pair of \emph{single}-layered dielectrically anisotropic slabs include a one-dimensional calculation~\cite{enk}, calculations on two ellipsoids with anisotropic dielectric function~\cite{mostepanenko}, a pair of dielectric slabs with different conductivity directions~\cite{kenneth}, and a quantum torque calculation for two specific uniaxial materials (barium titanate and quartz or calcite)~\cite{Capasso1,Capasso-erratum}. Experiments have also been proposed to measure the vdW torque using cholesteric liquid crystals~\cite{somers-munday}. Apart from the dielectric anisotropy, morphological anisotropy has been studied between anisotropic bodies \cite{Veble1,Emig-1,Emig-2} or even between surfaces that have anisotropic decorations \cite{Roya1,Roya2} and the effects of dielectric vs. morphological anisotropy have been delineated and compared \cite{hopkins}.

Dielectrically anisotropic \emph{multi}-layered materials have many examples, appearing in ceramics and clays, such as kaolinite~\cite{meunier} \bing{and computations of the vdW \emph{forces} for multi-layered systems are well-known in the literature~\cite{ninham-parsegian1,ninham-parsegian2,parsegian-ninham,php,pfp,Veble2,podgornik-parsegian2004,shao}}. \bing{In addition, many common minerals, e.g. micas, serpentine and chlorite to name a few, exist as different {\em polytypes} differing in layer-stacking configurations with repeated lateral offsets and rotations between the neighboring layers \cite{Polytypes}. These rotations of the layer orientations, implying also rotations in the principal axes of the respective dielectric tensor, implicate local long-range vdW \emph{torques} between the building blocks of the layered materials. These torques could play a stabilizing role favoring certain type of polytype with e.g. ordered periodic layer sequence as opposed to  random stacking sequences.} It is thus obviously important and relevant to investigate the vdW torques corresponding to such systems and the role it plays in the self-assembly \bing{and stabilization} of isolated single layers in crystallite structures, consisting of alternating dielectrically anisotropic crystal and isotropic (\bing{solvent}) layers.
From the nanoscale materials engineering side it is also of interest to create materials whose interactions can be tuned by suitable modifications of the internal structure~\cite{miguez1,miguez2}, \bing{as in the case of materials} composed of layers of different but known optical properties whose overall interaction behavior can be controlled by changing the thicknesses and the optical anisotropies of the individual layers.  

In the present Paper our objective is to investigate the vdW \emph{torque} as well as the interaction force between a pair of \emph{layered} slabs, each composed of coplanar layers that alternate between two distinct types of media: an optically anisotropic material and an isotropic \bing{(solvent)} material. Within each layered slab, the optic axis undergoes a constant angular increment across the anisotropic layers, \bing{while between} the slabs there is also a relative angular difference between the optic axes of the oppositely facing anisotropic layers. The way is thus paved to explore how the behaviors of the vdW torque and force change as one changes the following parameters: (i)~the thicknesses of the layers, (ii)~the angular \bing{dielectric} increment within each slab, and (iii)~the relative \bing{dielectric} angular difference between the slabs. Methodologically, our approach is a cross-pollination of the ideas and methods of previous approaches to determine the vdW torque between single-layered slabs and the transfer matrix method of computing the vdW force between multi-layered slabs~\cite{php,podgornik-parsegian2004}. We shall delimit ourselves to the \emph{non-retarded} limit, i.e., the limit where the speed of light is \bing{taken as} effectively infinite. This is a good approximation for slabs that are separated by distances smaller than $\sim$ 100 nm lengthscale. Furthermore, if the system is at high temperature (e.g., room temperature) and the intervening isotropic \bing{solvent} medium is water, at sufficiently large separations the zero Matsubara frequency term dominates over correction terms coming from retardation effects~\cite{parsegian-vdw}, and thus the non-retarded limit also provides a good approximation for the latter regime. 

\bing{In Sec.~II, we describe our system and develop a general formalism based on the method of transfer matrices. From Secs.~III to VI, we apply our formalism to illustrative, specific examples, in the simplifying approximation (of Ref.~\cite{parsegian-weiss}) that the dielectric anisotropy is weak and the dielectric susceptibility along the ordinary axes of the uniaxial layers matches the dielectric susceptibility of the solvent. We examine in turn a system with two single uniaxial layers, a single uniaxial layer interacting with a multilayered slab that has all its optic axes aligned, a single uniaxial layer interacting with a multilayered slab having rotating optic axes, and two interacting multilayered slabs each with rotating optic axes. In these systems the optic axes lie in the plane parallel to the layers. In Sec.~VII, we reconsider the previous systems but now with optic axes perpendicular to the plane of the layers.} 

\section{The system}
\begin{figure}
		\includegraphics[width=0.5\textwidth]{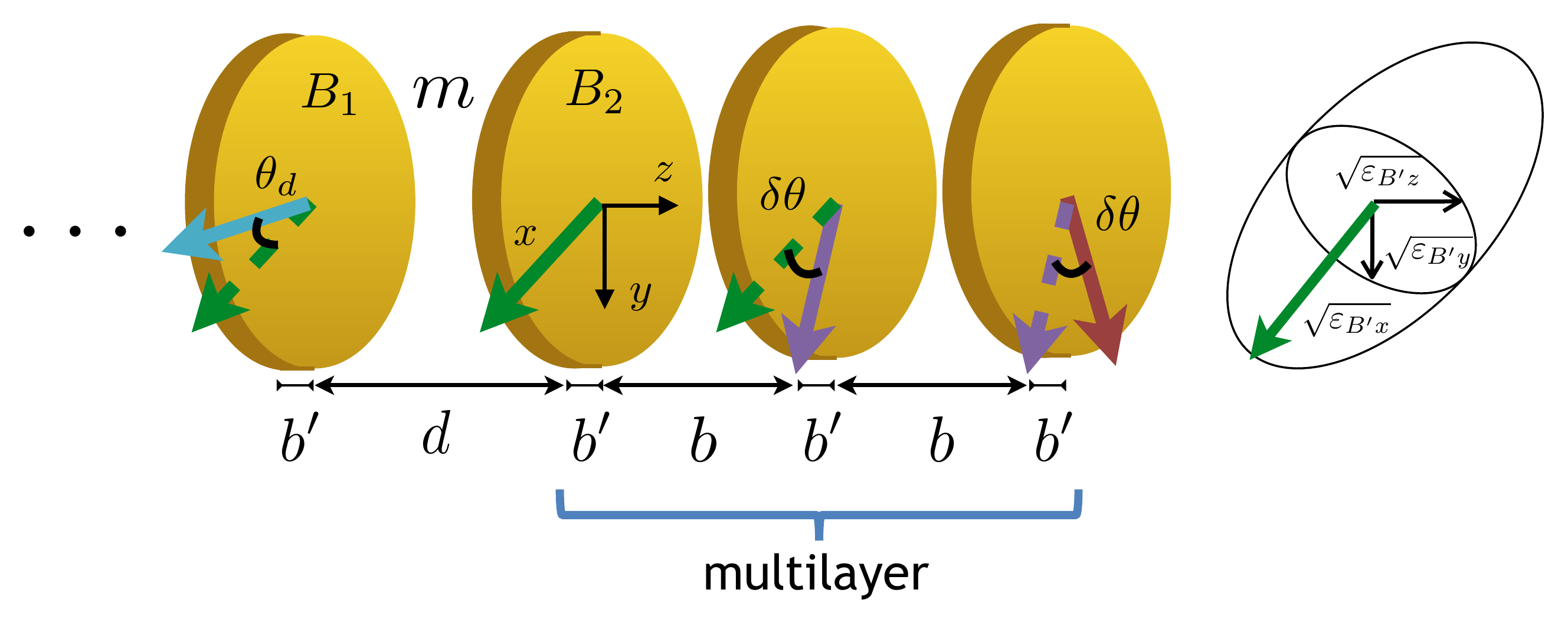}
	\caption{A model system of layered slabs. Each slab may consist of one or many \bing{{\em uniaxial}} optically anisotropic layers, each with the same thickness $b'$. Within each slab and between every pair of adjacent anisotropic layers is a layer of intervening isotropic \bing{(solvent)} medium of thickness $b$. The slabs are separated by a gap $m$ of width $d$, which has the same dielectric properties as the isotropic \bing{(solvent)} medium. The space coordinates have been chosen such that the $x$-axis is parallel to the optic axis (shown as the green unbroken arrow) of the left-most layer of the right slab (denoted by $B_2$, which we take to be the reference layer), and the optic axis of the right-most layer of the left slab has a relative angle $\theta_d$. The optic axis undergoes a constant rotation of $\delta\theta$ within each slab. Shown on the right is the dielectric ellipsoid corresponding to layer $B_2$. The \emph{geometric} and \emph{optical} anisotropies of the system should in no way be conflated, as the geometric axis points in the $z$-direction whereas the optic axes are perpendicular to $z$.}
\label{fig:anisotropy}
\end{figure}

Our model system (see Fig.~\ref{fig:anisotropy}) consists of a pair of co-axial and co-planar slabs with an intervening medium $m$ of thickness $d$. The \bing{(solvent)} medium $m$ is dielectrically \emph{isotropic} with dielectric permittivity $\varepsilon_m$. On the other hand, the slabs can either be single or multi-layered. The single-layered slab is dielectrically \emph{anisotropic}. In the multi-layered slab, there are $N+1$ dielectrically anisotropic layers (which we call type $B'$) and $N$ isotropic layers (which we call type $B$ \bing{-- this can be an aqueous or non-aqueous solvent, such as water or ethanol}), the layers alternating between dielectric anisotropy and isotropy. For example, the $B'$-type layer could represent silicate and the $B$-type layer could represent water in systems such as kaolinite clays~\cite{meunier}.  

\subsection{Dielectric tensor}
In terms of principal dielectric axes the dielectric tensor for each $B$-type layer is given by 
\be
\bm{\varepsilon}_B^{(\rm{prin})} = \varepsilon_{W} \mathbf{I}
\ee
where $\varepsilon_{W}$ is the dielectric permittivity of the isotropic medium (which, e.g., for water has a static value of $\sim 80 \varepsilon_0$ at $293 \, {\rm{K}}$, and $\varepsilon_0$ is the vacuum permittivity), and $\mathbf{I}$ is the identity matrix. Written in terms of principal axes, the dielectric tensor for the \emph{reference} $B'$-type layer is given by 
\be
\bm{\varepsilon}_{B'}^{(\rm{prin})} = 
\begin{pmatrix}
    \varepsilon_{B'x}      & 0 & 0 \\
    0 & \varepsilon_{B'y} & 0 \\
    0 & 0 & \varepsilon_{B'z}
\end{pmatrix}
\ee
where the $x$ and $y$ directions lie in the plane of the layer, and $z$ is perpendicular to the plane of the layer, see Fig. \ref{fig:anisotropy} (where layer $B_2$ is taken to be the reference layer).  Taking the dielectrically anisotropic material to be \emph{uniaxial} and defining the optic axis to be parallel to the $x$-axis, then $\varepsilon_{B'x} \neq \varepsilon_{B'y} = \varepsilon_{B'z}$. 
If layer $i$ is of type $B'$ (i.e., anisotropic) and its optic axis is rotated relative to the optic axis of the reference layer by an angle $\theta_i$, we can express the corresponding dielectric tensor as 
\ba
\label{eq:dielectric}
&&\bm{\varepsilon}^{(i)}(\theta_i) = 
\\
&&\!\!\!\!\!\!\!\!\!\!\!\!
\begin{pmatrix}
    \varepsilon_{B'x} \cos^2\theta_i +  \varepsilon_{B'y} \sin^2\theta_i & (\varepsilon_{B'x}-\varepsilon_{B'y})\sin\theta_i\cos\theta_i & 0 \\
    (\varepsilon_{B'x}-\varepsilon_{B'y})\sin\theta_i\cos\theta_i & \varepsilon_{B'x} \sin^2\theta_i +  \varepsilon_{B'y} \cos^2\theta_i & 0 \\
    0 & 0 & \varepsilon_{B'z}
\end{pmatrix}
\nonumber
\ea
If the $i$-th layer is type $B$ (i.e., isotropic) then $\bm{\varepsilon}^{(i)} = \varepsilon_W \mathbf{I}$. 

\subsection{van der Waals interaction free energy}
\label{sec:free_energy}

To calculate the free energy of vdW interaction we employ the van Kampen-Nijboer-Schram method~\cite{parsegian-vdw,kampen}, in which the electromagnetic field is represented as an ensemble of harmonic oscillators described by the Helmholtz free energy 
\be
F(T) = k_{\rm{B}} T \sum_{\{\omega_j\}} \ln (2\sinh(\beta \hbar \omega_j/2))
\ee
where $T$ is the temperature, $k_{\rm{B}}$ is Boltzmann's constant, $\beta=1/k_{\rm{B}} T$ is the inverse temperature, and $\hbar=h/2\pi$ where $h$ is Planck's constant. As the vdW interaction arises from correlations of electromagnetic surface fluctuational modes 
~\cite{sernelius}, the sum only include those mode frequencies $\omega_j$ that obey the dispersion relation, which is in general a nonlinear equation in $\omega_j$. The task of computing $F$ is however drastically simplified by the use of the Argument Principle~\cite{kampen,parsegian-vdw,mahanty}, via which the free energy can be transformed to the following more manageable form: 
\be
F(T) = k_{\rm{B}} T {\sum_{n=0}^{\infty}}^\prime \ln D(i\xi_n),
\label{eq:ft}
\ee
where the sum is over Matsubara frequencies, $\xi_n = (2\pi k_{\rm{B}}T/\hbar)n$, the prime denotes that we have to multiply the $n=0$ term by a factor  $1/2$, and \bing{$D(\omega_j)=0$} is the dispersion relation \bing{whose solutions are the mode frequencies $\omega_j$}.

\bing{In principle $D(\omega)$ is calculated from the full set of Maxwell equations, but here we delimit ourselves to the case of $c \rightarrow \infty$ which corresponds to the non-retarded case as discussed in detail in Refs.~\cite{kampen,parsegian-weiss}. (The retardation actually enters only for separations on the order of \bing{10-100} nm, which is not the case we are interested in.) The frequency summation comes from the poles of the $\ln(2 \sinh(\beta \omega/2))$, the free energy of the harmonic oscillators, and is not affected by the non-retardation approximation. We now turn to the evaluation of the dispersion relation.} 

\bing{
Here it may be also worth mentioning a distinction between (i)~the {\em non-retardation} limit, i.e. $c\rightarrow\infty$, which we have taken, and (ii)~the {\em zero frequency} limit, $\omega \rightarrow 0$. Both limits lead to vanishing right-hand sides in the Maxwell equations Eqs.~(\ref{eq:maxwell}). On the other hand, the zero frequency limit leads to a static value for the dielectric function, and thus a dispersion relation with no frequency dependence that can be reduced to (spatial) fluctuation determinant of the field modes \cite{Perspective}, whereas the dielectric function (and correspondingly the dispersion relation) retains its frequency dependence in the $c \rightarrow \infty$ limit. For details, see the discussion in e.g. Ref.~\cite{parsegian-vdw}}

\subsection{Dispersion relation}
The dispersion relation for the two-slab system can be derived from the boundary conditions that the electromagnetic surface modes have to satisfy. For a source- and current-free system, Maxwell's equations are given by 
\begin{subequations}
\label{eq:maxwell}
\ba
\nabla\times\mathbf{H} &=& \frac{1}{c} \frac{\partial \mathbf{D}}{\partial t}, \quad \nabla\cdot\mathbf{D} = 0, 
\\
-\nabla \times \mathbf{E} &=& \frac{1}{c} \frac{\partial \mathbf{B}}{\partial t}, \quad \nabla\cdot\mathbf{B} = 0 
\ea
\end{subequations}
where $\mathbf{D} = \bm{\varepsilon} \cdot \mathbf{E}$ and $\mathbf{B} = \bm{\mu} \cdot\mathbf{H}$, where $\bm{\varepsilon}$ and $\bm{\mu}$ are the dielectric and magnetic tensors. We assume that the dielectric properties are anisotropic but the magnetic properties are isotropic, so we will set $\bm{\mu} = \mathbf{I}$, where $\mathbf{I}$ is the unit matrix. For the case of slab geometry, it is known that there are two sets of solutions to Maxwell's equations, viz., the TM and TE modes, describing the two different polarizations of the electromagnetic wave. In what follows, we shall consider only the TM mode contribution to the vdW free energy and neglect the TE mode contribution, as the effect of dielectric anisotropy is present only in \bing{the former~\cite{footnote:tm}}. 

In the \emph{non-retarded} regime, $c \rightarrow \infty$, and the equations governing the electric field become
\begin{subequations}
\ba
\label{eq:efield}
&&\partial_a (\varepsilon_{ab} \rm{E}_b) = 0,
\\
&&\epsilon_{abc}\partial_b \rm{E}_c = 0.
\ea
\end{subequations}
Here $a,b,c=1,2,3$ (or $x,y,z$) are Cartesian indices labeling the directions in space, $\epsilon_{abc}$ is the completely antisymmetric tensor, $\partial_a \equiv \partial/\partial x_a$ where $x_1=x$, $x_2=y$ and $x_3=z$. Solving these equations subject to boundary conditions in the slab geometry leads to the TM mode. 
Owing to the curl-free condition we can also represent the electric field as the gradient of a scalar potential $\varphi$: 
\be
\mathbf{E} = -\nabla\varphi, 
\ee
whence we obtain 
\be
\partial_a (\varepsilon_{ab} \partial_b \varphi) = 0.
\label{eq:maxwell1}
\ee
This has to be solved with respect to the boundary conditions that both $\varphi$ and $(\bm{\varepsilon}\!\cdot\!\mathbf{E})_z$ are continuous across the interface between every pair of adjacent layers. As the translational symmetry is broken along the $3$-direction we can represent $\varphi$ in terms of a two-dimensional Fourier transform:
\be
\varphi_i(\xv_\perp, z) = \int\frac{du\,dv}{(2\pi)^2} e^{i(ux+vy)} f_{i}(z),
\ee
where $\xv_\perp=(x,y)$, $u, v$ are the momenta in the $x$ and $y$ directions, and the subscript $i$ labels the \bing{layer~\cite{kampen}}. Plugging this into Eq.~(\ref{eq:maxwell1}) gives
\be
\partial_z^2 f_i (z) - \rho_i^2(\theta_i) f_i(z) = 0.
\label{eq:f}
\ee
If layer $i$ is an \emph{isotropic} medium then $\rho_i = \sqrt{u^2+v^2}$. On the other hand, if layer $i$ is dielectrically \emph{anisotropic} with its optic axis lying in the plane of the layer, then
\ba
\label{eq:rho_compliquee}
\rho_i^2(\theta_i) 
&\equiv& 
\frac{\varepsilon_{11}^{(i)} u^2 + 2\varepsilon_{12}^{(i)} uv + \varepsilon_{22}^{(i)} v^2}{\varepsilon_{B'z}}
\\
&=& 
\frac{\varepsilon_{B'x}}{\varepsilon_{B'z}}(u \cos\theta_i + v\sin\theta_i)^2 
\nonumber\\
&&+ 
\frac{\varepsilon_{B'y}}{\varepsilon_{B'z}}(v \cos\theta_i - u\sin\theta_i)^2, 
\nonumber
\ea
where $\varepsilon_{ab}^{(i)}$ denotes the $ab$ element of the dielectric tensor in Eq.~(\ref{eq:dielectric}). The solution to Eq.~(\ref{eq:f}) is given by 
\be
f_i(z) = A_{i} e^{\rho_i z} + B_{i} e^{-\rho_i z}. 
\ee
Continuity of $\varphi$ and $(\bm{\varepsilon} \cdot \mathbf{E})_z$ at the interface between layers $i$ and $i+1$ demands
\ba
f_{i+1}(\ell_{i,i+1}) &=& f_i(\ell_{i,i+1}),
\\
\varepsilon_{33}^{(i+1)} \partial_z f_{i+1}(\ell_{i,i+1}) &=& \varepsilon_{33}^{(i)} \partial_z f_i(\ell_{i,i+1}) 
\ea
where $z=\ell_{i,i+1}$ is the position of the interface. These equations lead to 
\begin{widetext}
\be
\label{eq:transmit}
\begin{pmatrix}
A_{i+1} \\ B_{i+1}
\end{pmatrix}
= 
-\frac{1}{2} 
\bigg( 1+\frac{\varepsilon_{33}^{(i)} \rho_{i}}{\varepsilon_{33}^{(i+1)} \rho_{i+1}} \bigg)
\begin{pmatrix}
e^{-(\rho_{i+1}-\rho_i)\ell_{i,i+1}} & \bar{\Delta}_{i+1,i} e^{-(\rho_{i+1}+\rho_i)\ell_{i,i+1}} \\
\bar{\Delta}_{i+1,i} e^{(\rho_{i+1}+\rho_i)\ell_{i,i+1}} & e^{(\rho_{i+1}-\rho_i)\ell_{i,i+1}} 
\end{pmatrix}
\begin{pmatrix}
A_i \\ B_i
\end{pmatrix},
\ee
\end{widetext}
where we have defined a reflection coefficient describing the dielectric discontinuity across the interface between layers $i$ and $i+1$: 
\be
\bar{\Delta}_{i+1,i} \equiv \frac{\varepsilon_{33}^{(i+1)} \rho_{i+1}-\varepsilon_{33}^{(i)} \rho_{i}}{\varepsilon_{33}^{(i+1)} \rho_{i+1}+\varepsilon_{33}^{(i)} \rho_{i}}. 
\label{eq:Delta}
\ee
Denoting the left-most layer of the slab on the left by the index $L$, and the right-most layer of the slab on the right by the index $R$, the dispersion relation is obtained from the condition that $A_R = B_L = 0$. We are thus at liberty to ignore the prefactor on the right-hand side (RHS) of Eq.~(\ref{eq:transmit}) and redefine amplitudes such that 
\ba
\begin{pmatrix}
\tA_{i+1} \\ \tB_{i+1}
\end{pmatrix} 
&\!\!=\!\!& 
\begin{pmatrix}
e^{\rho_{i+1}\ell_{i,i+1}} & 0 \\ 
0 & e^{-\rho_{i+1}\ell_{i,i+1}}
\end{pmatrix} 
\begin{pmatrix}
A_{i+1} \\ B_{i+1}
\end{pmatrix} 
\\
\begin{pmatrix}
\tA_{i} \\ \tB_{i}
\end{pmatrix} 
&\!\!=\!\!& 
\begin{pmatrix}
e^{\rho_{i}\ell_{i-1,i}} & 0 \\ 
0 & e^{-\rho_{i}\ell_{i-1,i}}
\end{pmatrix} 
\begin{pmatrix}
A_{i} \\ B_{i}
\end{pmatrix}.
\ea
We can write
\be
\begin{pmatrix}
\tA_{i+1} \\ \tB_{i+1}
\end{pmatrix} = 
e^{\rho_{i}(\ell_{i,i+1}-\ell_{i-1,i})}
\mathbf{M}_{i+1,i}\cdot
\begin{pmatrix}
\tA_{i} \\ \tB_{i}
\end{pmatrix}
\label{eq:rela}
\ee
where 
\be
\mathbf{M}_{i+1,i} \equiv 
\begin{pmatrix}
1 & -\bar{\Delta}_{i,i+1} e^{-2\rho_i(\ell_{i,i+1}-\ell_{i-1,i})} \\ 
-\bar{\Delta}_{i,i+1} & e^{-2\rho_i(\ell_{i,i+1}-\ell_{i-1,i})}
\end{pmatrix}
\ee
We can further decompose $\mathbf{M}_{i+1,i}$ into the product of two matrices: 
\be
\mathbf{M}_{i+1,i} = \mathbf{D}_{i+1,i} \cdot \mathbf{T}_{i},
\ee
where 
\ba
\mathbf{D}_{i+1,i} &\equiv& 
\begin{pmatrix}
1 & -\bar{\Delta}_{i+1,i} \\
-\bar{\Delta}_{i+1,i} & 1
\end{pmatrix},
\\
\mathbf{T}_{i} 
&\equiv& 
\begin{pmatrix}
1 & 0 \\
0 & e^{-2\rho_i (\ell_{i+1,i}-\ell_{i,i-1})}
\end{pmatrix},
\label{eq:Tmatrix}
\ea
We can practically ignore the prefactor $e^{\rho_{i}(\ell_{i,i+1}-\ell_{i-1,i})}$ in subsequent calculations because it does not affect the dispersion relation. By induction we can relate the coefficients $\tA_R$ and $\tB_R$ of the right-most layer to the coefficients $\tA_L$ and $\tB_L (=0)$ of the left-most layer, viz., 
\be
\begin{pmatrix}
\tA_{R} \\ \tB_{R}
\end{pmatrix} = 
\mathbf{\Theta} \cdot
\begin{pmatrix}
\tA_{L} \\ 0
\end{pmatrix},
\ee
where the overall transfer matrix $\mathbf{\Theta}$ is given by 
\be
\mathbf{\Theta} \equiv \mathbf{D}_{R,P} \prod_{i=0}^{\bing{N}-1} \mathbf{T}_{i+1} \mathbf{D}_{i+1,i}, 
\ee
$L$ corresponds to the $i=0$ layer, and we have assumed that there is a total of $\bing{N}+1$ layers in the system, of which the end layers on the left and the right are semi-infinite. If we consider the effective interaction between two layered slabs separated by a gap of isotropic medium of width $d$, then the dispersion relation is given by
\be
D(d,\omega) = \frac{\Theta_{11}(d,\omega)}{\Theta_{11}(d\rightarrow\infty,\omega)} = 0,
\label{nkjfsd}
\ee
where $\Theta_{11}$ is the $11$ component of the transfer matrix. This follows since both media (L) and (R) are semi-infinite and the fields should decay far away from the dielectric boundaries, and thus $\tA_{R} = 0$ and $\tB_{L} = 0$. This can only happen if $\Theta_{11} \equiv 0$, i.e., if Eq.~(\ref{nkjfsd}) is valid. \bing{In Eq.~(\ref{nkjfsd}) we have normalized the dispersion relation by its value for infinitely separated layers. According to the definition of the vdW free energy, Eq.~(\ref{eq:ft}), this amounts to the same thing as subtracting the bulk contribution from the complete free energy, with the remainder obviously being just the vdW {\em interaction free energy}.}

\subsection{Anisotropy factor}
By writing 
\be
u = Q \cos\psi, \, v = Q \sin \psi,
\ee 
we can rewrite $\rho_i$ (cf. Eq.~(\ref{eq:rho_compliquee})) in the simpler form:
\be
\rho_i = Q \, g_i (\theta_i-\psi), 
\ee
where the effects of dielectric anisotropic are now contained inside the anisotropy factor $g_i$. 
For isotropic, $B$-type media, $g_i = 1$, whilst for anisotropic, $B'$-type media, it is given by 
\be
g_i(\theta_i-\psi) \equiv \sqrt{\frac{\varepsilon_{B'y}}{\varepsilon_{B'z}} + \frac{\varepsilon_{B'x}-\varepsilon_{B'y}}{\varepsilon_{B'z}} \cos^2(\theta_i-\psi)}
\ee 
The two-dimensional integral measure becomes 
\be
du \, dv = Q \, dQ \, d\psi.
\ee
The preceding formal discussions will be fleshed out more fully in the following sections, where we apply our formalism to concrete examples. 

\section{Two interacting single anisotropic layers} 


\label{sec:singlelayers}
\begin{figure}
		\includegraphics[width=0.3\textwidth]{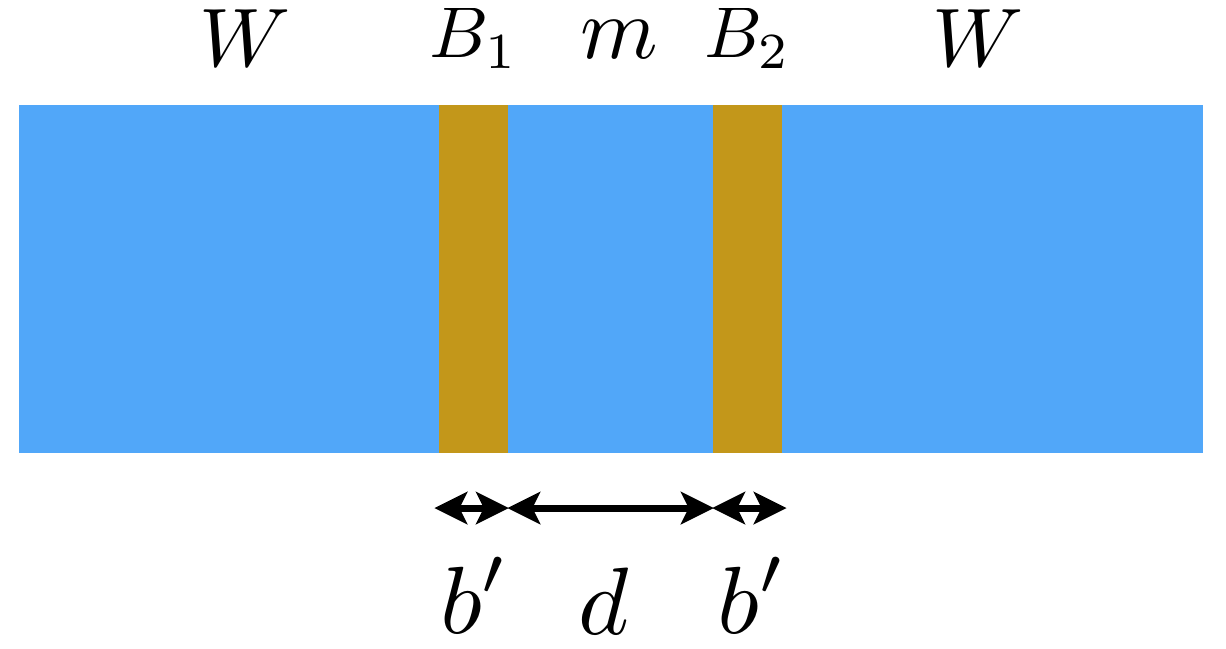}
	\caption{A pair of single anisotropic layers $B_1$ and $B_2$ of the same thickness $b'$ interacting across an intervening isotropic solvent medium $m$ of thickness $d$. The layers $B_1$ and $B_2$ are also bounded on the left and the right respectively by \bing{the same solvent} $W$.}
\label{fig:system_ss}
\end{figure}
We consider two co-axial parallel single layers $B_1$ and $B_2$ composed of an anisotropic material (for example, silicate), each being of thickness $b'$, separated by an intervening isotropic medium (for example water) of width $d$ and dielectric permittivity $\varepsilon_W$, and the media to the left of $B_1$ and the right of $B_2$ are also isotropic and of the same dielectric permittivity $\varepsilon_W$ (see Fig.~\ref{fig:system_ss}). The optic axis of $B_2$ is however rotated relative to the optic axis of $B_1$ by an angle $\theta_{B_2}-\theta_{B_1}$. Using transfer matrices, we can express this set-up by 
\be
\begin{pmatrix}
\bing{\widetilde{A}_{R}} \\ \bing{\widetilde{B}_{R}}
\end{pmatrix} = 
\mathbf{\Theta}^{(\rm{ss})} \cdot
\begin{pmatrix}
\bing{\widetilde{A}_{L}} \\ 0
\end{pmatrix}
\ee
where the overall transfer matrix is given by 
\be
\label{eq:Theta_ss}
\mathbf{\Theta}^{(\rm{ss})} \equiv \mathbf{D}_{WB_1}  \mathbf{T}_{B_1} 
\mathbf{D}_{B_1W}  \mathbf{T}_{m} \mathbf{D}_{WB_2} \mathbf{T}_{B_2} \mathbf{D}_{B_2W} 
\ee
and the boundary condition that $A_R=0$ can be enforced via the requirement that $\Theta_{11}^{(\rm{ss})}=0$. The matrices are given by 
\ba
\mathbf{D}_{WB_1} 
&\equiv&
\begin{pmatrix}
1 & -\bar{\Delta}_{WB_1} \\
-\bar{\Delta}_{WB_1} & 1
\end{pmatrix},
\\
\mathbf{D}_{WB_2} 
&\equiv&
\begin{pmatrix}
1 & -\bar{\Delta}_{WB_2} \\
-\bar{\Delta}_{WB_2} & 1
\end{pmatrix},
\\
\mathbf{T}_{B_1} 
&\equiv& 
\begin{pmatrix}
1 & 0 \\
0 & e^{-2 Q b' g_{B_1}}
\end{pmatrix}, 
\\
\mathbf{T}_{B_2} 
&\equiv& 
\begin{pmatrix}
1 & 0 \\
0 & e^{-2 Q b' g_{B_2}}
\end{pmatrix}, 
\\
\mathbf{T}_{m} 
&\equiv& 
\begin{pmatrix}
1 & 0 \\
0 & e^{-2 Q d}
\end{pmatrix}, 
\ea
and $\mathbf{D}_{B_1 W}$ ($\mathbf{D}_{B_2 W}$) corresponds to $\mathbf{D}_{W B_1}$ ($\mathbf{D}_{W B_2}$) with $\bar{\Delta}_{WB_1}$ ($\bar{\Delta}_{WB_2}$) replaced by $-\bar{\Delta}_{WB_1}$ ($-\bar{\Delta}_{WB_2}$).
The reflection coefficients are given by 
\ba
\label{eq:Deltas}
\bar{\Delta}_{WB_1} &\equiv& \frac{1 - g_{B_1}}{1 + g_{B_1}}, 
\,
\bar{\Delta}_{WB_2} \equiv \frac{1 - g_{B_2}}{1 + g_{B_2}}, 
\\
\bar{\Delta}_{WB_1} &=& -\bar{\Delta}_{B_1W}, 
\, \bar{\Delta}_{WB_2} = -\bar{\Delta}_{B_2W}. 
\ea
We assume that the anisotropic layers have the same dielectric properties (apart from the orientation of the optic axis) and are \emph{uniaxial} (i.e., the dielectric property along the optic axis is different from the dielectric properties along the other two principal axes, and the dielectric properties along the latter axes are identical). 

\bing{In addition, following Ref.~\cite{parsegian-weiss}, we adopt the simplifying assumption that the dielectric permittivity along each of the non-optic principal (i.e., {\em ordinary}) axes is equal to the dielectric permittivity of the isotropic media: $\varepsilon_{B_1,y}=\varepsilon_{B_1,z}=\varepsilon_{B_2,y}=\varepsilon_{B_2,z}=\varepsilon_{W}$ and $\varepsilon_{B_1,x}=\varepsilon_{B_2,x}$. A possible realization where such dielectric matching between the ordinary axes of the anisotropic layer and the solvent holds approximately is a stack of ${\rm LiNbO_3}$ layers immersed in water at room temperature~\cite{footnote:matching}; both static dielectric constants of the solvent and the anisotropic layer along the ordinary axes are approximately 80 at room temperature~\cite{young-frederikse}.} 

Let us also define a quantity that characterizes the anisotropy between the principal dielectric permittivities, viz.,  
\be
\gamma_n \equiv \varepsilon_{B_1,x}(i\xi_n)/\varepsilon_{B_1,z}(i\xi_n) - 1,
\ee 
where the subscript $n$ reflects the dependence of the dielectric anisotropy on the frequency. 
The anisotropy factors are then expressible by 
\begin{subequations}
\ba
g_{B_1} &=& \sqrt{1 + \gamma_n(\cos(\theta_{B_1}-\psi))^2}, 
\\
g_{B_2} &=& \sqrt{1 + \gamma_n(\cos(\theta_{B_2}-\psi))^2}. 
\ea
\end{subequations}
By using the matrices above, we can readily compute the 11 element of the overall transfer matrix. Its value is given in Eq.~(\ref{eq:Theta11_ss}) of App.~\ref{app:eli}. 

\subsection{Interaction between isotropic layers}
As a check of consistency, let us consider layers that are dielectrically \emph{isotropic} (i.e., $g_{B_1}=g_{B_2}=1$) and $\bar{\Delta}_{W B_1} = \bar{\Delta}_{W B_2} = \bar{\Delta}_{W B_1} = \bar{\Delta}_{W B_2} \equiv \bar{\Delta}$; in this case, we have (using Eq.~(\ref{eq:Theta11_ss})) 
\be
\Theta_{11}^{({\rm ss})} (d,\bing{\omega}) = (1 - \bar{\Delta}^2 e^{-2Q b'})^2 - \bar{\Delta}^2 e^{-2Q d} (1 - e^{-2Q b'})^2
\label{eq:T11}
\ee
\bing{Using Eqs.~(\ref{eq:ft}) and (\ref{eq:T11}), we find that the} \bing{interaction} free energy per unit area is given by 
\ba
\label{eq:G_simple}
G_{ss} &=& \frac{k_{\rm{B}}T}{2\pi} \! \sum_n^\prime \! \int_0^\infty \!\!\! dQ \, Q
\ln \frac{\Theta_{11}^{({\rm ss})} (d, \bing{i\xi_n})}{\Theta_{11}^{({\rm ss})} (d\rightarrow\infty, \bing{i\xi_n})} 
\\
&=&
 \frac{k_{\rm{B}}T}{2\pi} \! \sum_n^\prime \! \int_0^\infty \!\!\! dQ \, Q
\ln \bigg[ 1 - \frac{\bar{\Delta}^2 e^{-2Q d} (1 - e^{-2Q b'})^2}{(1 - \bar{\Delta}^2 e^{-2Q b'})^2} \bigg].
\nonumber 
\ea
This agrees with previous results~\cite{parsegian-vdw,mahanty,pfp} on an interacting pair of dielectrically isotropic layers immersed in an isotropic solvent. 

\subsection{Interaction between anisotropic layers}

We return to Eq.~(\ref{eq:Theta11_ss}) and consider weak anisotropy, for which $\gamma_n \ll 1$, \bing{as is the case in materials with weak dielectric anisotropies that include amongst others the calcite, whose static dielectric susceptibility along the ordinary axes is 8.5 and that along the optic axis is 8 at room temperature~\cite{Capasso1}.} To leading order we have 
\ba
&&\Theta_{11}^{(\rm{ss})} \approx 1 - 
\frac{\gamma_n^2}{16}e^{-2Qb'} 
\nonumber\\
&&\times
\big(  
4e^{-2Qd} \sinh^2(Qb') \cos^2(\theta_{B_1}-\psi) \cos^2(\theta_{B_2}-\psi) 
\nonumber\\
&&+ \cos^4(\theta_{B_1}-\psi) + \cos^4(\theta_{B_2}-\psi) 
\big)
\ea
The corresponding \bing{interaction} free energy per unit area is given by 
\be
\label{eq:Gss}
G_{ss} = \frac{k_{\rm{B}}T}{4\pi^2} \! \sum_n^\prime \! \int du\,dv\,\ln \frac{\Theta_{11}^{(\rm{ss})}(d)}{\Theta_{11}^{(\rm{ss})}(d\rightarrow\infty)}
\ee
For weak anisotropy, this leads to 
\ba
G_{ss} 
&\approx& 
-\frac{\gamma^{2} k_{\rm{B}}T}{32\pi^2} \! 
\int_0^{2\pi} \!\!\!\! d\psi \!\int_0^\infty \!\!\!\!dQ \, Q \, 
e^{-2Q(b'+d)} \sinh^2(Qb') 
\nonumber\\
&&\quad\times
\cos^2(\theta_{B_1}-\psi) \cos^2(\theta_{B_2}-\psi)
\nonumber\\
&=& 
-\frac{\gamma^{2} k_{\rm{B}}T}{2048\pi} (1+2\cos^2(\theta_{B_1}-\theta_{B_2})) 
\nonumber\\
&&\quad\times
\left[ 
\frac{1}{d^2} - \frac{2}{(d+b')^2} + \frac{1}{(d+2b')^2}
\right],
\label{eq:Gss_app}
\ea
where we have defined $\gamma^2 \equiv 2\sum_n^\prime \gamma_n^2$. 
For \emph{large} separation ($d \gg b'$), we have 
\be
\label{eq:Gss_far}
G_{ss} 
\approx 
-\frac{3\gamma^{2} k_{\rm{B}}T (1+2\cos^2(\theta_{B_1}-\theta_{B_2})) (b')^2}{1024 \pi d^4}, 
\ee
which corresponds to an attractive force per unit area that decays with $d^{-5}$, viz., 
\be
\mathcal{F}_{ss} 
\approx 
-\frac{3\gamma^{2} k_{\rm{B}}T (1+2\cos^2(\theta_{B_1}-\theta_{B_2})) (b')^2}{256 \pi d^5}. 
\ee
For \emph{small} separation ($d \ll b'$), we have 
\be
G_{ss} 
\approx 
-\frac{\gamma^{2} k_{\rm{B}}T (1+2\cos^2(\theta_{B_1}-\theta_{B_2}))}{2048\pi d^2},
\label{eq:Gss_near}
\ee
which corresponds to an attractive force per unit area that decays with $d^{-3}$, viz., 
\be
\mathcal{F}_{ss} 
\approx 
-\frac{\gamma^{2} k_{\rm{B}}T (1+2\cos^2(\theta_{B_1}-\theta_{B_2}))}{1024\pi d^3}.
\ee
This is in agreement with the result obtained by Parsegian and Weiss~\cite{parsegian-weiss} in the case of non-retarded interaction. This agreement comes about because two layers that are separated by a distance much smaller than their thicknesses effectively resemble a pair of \emph{thick} slabs. Also, notably, the $\cos^2(\theta_{B_1}-\theta_{B_2})$ dependence on the relative anisotropy angle change, signals the material dielectric anisotropy effect~\cite{hopkins}. 

\subsection{van der Waals torque}
We can now straightforwardly derive the vdW torque per unit area $\tau$, by applying the general definition  
\be
\tau = -\frac{\partial G}{\partial \theta_d},
\ee
where $\theta_d \equiv \theta_{B_1} - \theta_{B_2}$. 
We consider the weak anisotropy regime and multi-layered slabs with a large number of $B'$-type layers. From Eq.~(\ref{eq:Gss_app}) we obtain the torque per unit area for two interacting single layered slabs, $\tau_{ss}$:
\be
\tau_{ss} = -\frac{\gamma^2 k_{{\rm B}}T \sin 2\theta_d}{1024\pi} \left[ \frac{1}{d^2} - \frac{2}{(d+b')^2} + \frac{1}{(d+2b')^2} \right]. 
\label{eq:tss}
\ee
For $d \gg b'$, the torque is approximately given by 
\be
\tau_{ss} \approx -\frac{3\gamma^2 k_{{\rm B}}T (b')^2}{512\pi d^4} \sin(2\theta_d),
\ee
while for $d \ll b'$, the torque is approximated by 
\be
\tau_{ss} \approx -\frac{\gamma^2 k_{{\rm B}}T}{1024\pi d^2} \sin(2\theta_d).
\ee
This latter limit is the same as that obtained by Parsegian and Weiss~\cite{parsegian-weiss} in the limit of non-retardation, as two single layers separated by a distance much smaller than their individual thicknesses is approximately equivalent to two thick slabs. 

Thus for both the vdW interaction free energy and torque, there is a crossover in the scaling behavior with separation from $d^{-2}$ to $d^{-4}$ as the separation increases beyond a lengthscale set by the thickness of each anisotropic layer. The vdW \emph{force} is always attractive, owing to $G_{ss}$ being always negative (and growing in magnitude as the separation decreases). On the other hand, the vdW \emph{torque} can change sign depending on $\theta_d$. For $0 < \theta_d < \pi/2$ and $\pi < \theta_d < 3\pi/2$, $\tau_{ss} < 0$, which implies that the configuration in which the optic axes of the two anisotropic layers are \emph{aligned} (or anti-aligned) is \emph{stable}: any deviation from alignment will generate an attractive torque that tends to restore the two layers to the aligned configuration. For $\pi/2 < \theta_d < \pi$ and $3\pi/2 < \theta_d < 2\pi$, $\tau_{ss} > 0$, which implies that the configuration in which the optic axes are \emph{perpendicular} is \emph{unstable}, as a slight deviation will bring about a repulsive torque that drives the layers away from their initial angular configuration. 

\section{Single anisotropic layer interacting with multilayer having aligned optic axes}
\label{sec:sm}

\begin{figure}
		\includegraphics[width=0.48\textwidth]{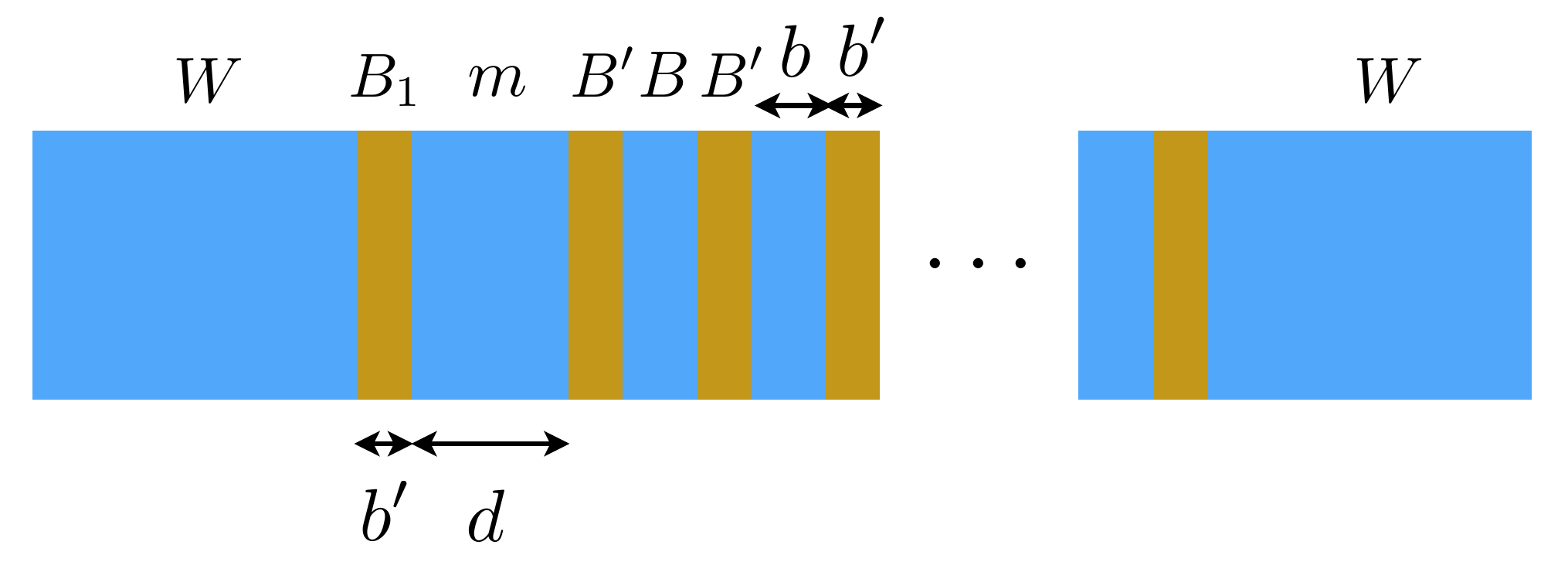}
	\caption{A single anisotropic layer $B_1$ of thickness $b'$ interacts with a slab composed of a sequence of alternating $B'$-type (anisotropic) and $B$-type (isotropic) layers, of thicknesses $b'$ and $b$ respectively, across an intervening isotropic medium $m$ of thickness $d$. The layer $B_1$ and the slab are bounded on the left and the right respectively by isotropic media $W$.}
\label{fig:system_sm}
\end{figure} 
Next, we consider a set-up in which we have a semi-infinite slab of the isotropic $B$-type medium, bounded on the right by a single layer $B_1$ made of anisotropic medium $B'$ (of thickness $b'$), followed by a gap $m$ of width $d$ which is composed of an isotropic $B$-type medium, and this is followed in turn by $N$ repeats of the $B'$-type layer (of thickness $b'$) and $B$-type layer (of thickness $b$), with a final $B'$ layer that is followed by a semi-infinite slab of the $B$-type medium (see Fig.~\ref{fig:system_sm}). 
The optic axis of the $B'$-type single layer on the left is oriented at an angle $\theta_d$ with respect to the optic axis of the left-most $B'$-type layer of the multilayered slab (the latter axis being our reference axis), and the optic axis of every $B'$-type layer inside the multilayered slab has the same orientation. 
Mathematically, this set-up is represented by the sequence of transfer matrices: 
\be
\begin{pmatrix}
\bing{\widetilde{A}_{R}} \\ \bing{\widetilde{B}_{R}}
\end{pmatrix} = 
\mathbf{\Theta}^{(\rm{sm})} \cdot
\begin{pmatrix}
\bing{\widetilde{A}_{L}} \\ 0
\end{pmatrix}
\ee
where $\mathbf{\Theta}^{(\rm{sm})}$ is given by 
\be
\label{eq:Theta_sm}
\mathbf{\Theta}^{(\rm{sm})} 
\equiv
\mathbf{A}^{N} 
\mathbf{D}_{WB'} \mathbf{T}_{B'} \mathbf{D}_{B'W} \mathbf{T}_{m} \mathbf{D}_{W B_1} 
 \mathbf{T}_{B_1} \mathbf{D}_{B_1 W}.
\ee
Here the matrix $\mathbf{A}$ describes a bilayer consisting of a $B$-type layer and a $B'$-type layer: 
\be
\label{eq:A}
\mathbf{A} \equiv \mathbf{D}_{WB'}  \mathbf{T}_{B'} \mathbf{D}_{B'W} \mathbf{T}_{B}.
\ee
Here $\mathbf{D}_{WB'}$ and $\mathbf{D}_{W B_1}$ describe the dielectric discontinuity at the interfaces between the $B$ and $B'$-type layers, and are given by
\ba
\mathbf{D}_{WB'} 
&\equiv&
\begin{pmatrix}
1 & -\bar{\Delta}_{WB'} \\
-\bar{\Delta}_{WB'} & 1
\end{pmatrix},
\\
\label{eq:DWL}
\mathbf{D}_{W B_1} 
&\equiv&
\begin{pmatrix}
1 & -\bar{\Delta}_{W B_1} \\
-\bar{\Delta}_{W B_1} & 1
\end{pmatrix},
\\
\bar{\Delta}_{W B_1} &\equiv& \frac{1 - g_{B_1}(\theta_d-\psi)}{1 + g_{B_1}(\theta_d-\psi)} = -\bar{\Delta}_{B_1 W}, 
\\
\bar{\Delta}_{WB'} &\equiv& \frac{1 - g_{B'}(\psi)}{1 + g_{B'}(\psi)} = -\bar{\Delta}_{B'W}, 
\ea
where $g_{B_1}$ and $g_{B'}$ are anisotropy factors. For a system in which $\varepsilon_{B'y}=\varepsilon_{B'z}=\varepsilon_{W}$, this is given by
\begin{subequations}
\ba
g_{B_1} &\equiv& \sqrt{1+\gamma_n \cos^2 (\theta_d - \psi)}, 
\\
g_{B'} &\equiv& \sqrt{1+\gamma_n \cos^2 \psi},
\ea
\end{subequations}
where we have defined 
\be
\gamma_n \equiv \varepsilon_{B'x}(i\xi_n)/\varepsilon_{B'y}(i\xi_n) - 1.
\ee
The matrices $\mathbf{T}_{B}$ and $\mathbf{T}_{B'}$ are related to the thicknesses of the $B$-type and $B'$-type layers respectively, and are given by
\ba
\mathbf{T}_{B} 
&\equiv& 
\begin{pmatrix}
1 & 0 \\
0 & e^{-2 Q b}
\end{pmatrix}, 
\\
\mathbf{T}_{B'} 
&\equiv& 
\begin{pmatrix}
1 & 0 \\
0 & e^{-2 Q b'  g_{B'}}
\end{pmatrix}
\ea
Similarly $\mathbf{T}_{B_1}$ and $\mathbf{T}_{m}$ are related to the thicknesses of the layer $B_1$ and the gap $m$:
\ba
\label{eq:TL}
\mathbf{T}_{B_1} 
&\equiv& 
\begin{pmatrix}
1 & 0 \\
0 & e^{-2 Q b' g_{B_1}}
\end{pmatrix}, 
\\
\label{eq:Tm}
\mathbf{T}_{m} 
&\equiv& 
\begin{pmatrix}
1 & 0 \\
0 & e^{-2 Q d}
\end{pmatrix}, 
\ea
where
\be
g_{B_1} \equiv \sqrt{1+\gamma_n \cos^2(\theta_d-\psi)}.
\label{eq:gL}
\ee
As before the boundary condition that $A_R=0$ can be enforced via the dispersion relation: $\Theta_{11}^{(\rm{sm})}=0$. The elements of the matrix $\mathbf{A}$ are given by Eqs.~(\ref{eq:Asmele}) of App.~\ref{app:1}. 

The matrix product $\mathbf{A}^N$ can be found using Abel\`{e}s' formula~(see, e.g., \bing{Ref.~\cite{born-wolf}}), which gives
\be
\label{eq:AN_matrix}
\mathbf{A}^{N} = 
\begin{pmatrix}
A_{11}^{(N)} & A_{12}^{(N)} \\
A_{21}^{(N)} & A_{22}^{(N)}
\end{pmatrix}
\ee
where 
\begin{subequations}
\label{eq:AN_matrix_coeffs}
\ba
A_{11}^{(N)} &\equiv& \bigg( \frac{A_{11}}{\sqrt{|\mathbf{A}|}} U_{N-1} - U_{N-2} \bigg) 
|\mathbf{A}_n|^{N/2}
\\
A_{12}^{(N)} &\equiv& A_{12} U_{N-1} |\mathbf{A}|^{(N-1)/2}
\\
A_{21}^{(N)} &\equiv& A_{21} U_{N-1} |\mathbf{A}|^{(N-1)/2}
\\
A_{22}^{(N)} &\equiv& \bigg( \frac{A_{22}}{\sqrt{|\mathbf{A}|}} U_{N-1} - U_{N-2} \bigg) 
|\mathbf{A}|^{N/2}
\ea
\end{subequations}
Here $U_N$ are the Chebyshev polynomials, with $U_0=1$ and $U_N=0$ for $N < 0$, and for $N > 0$ they are given by
\be
\label{eq:chebyshev}
U_{N-1} = \frac{\sinh N \xi}{\sinh \xi}
\ee
where 
\be
\label{eq:xi_def}
\xi \equiv \cosh^{-1} \bigg( \frac{A_{11}+A_{22}}{2\sqrt{|\mathbf{A}|}} \bigg).
\ee
The above representation for Chebyshev polynomials is valid for $\xi > 1$, which is the case as can be verified easily. For weak anisotropy, we can approximate $\xi$ by 
\be
\xi \approx Q(b+b') + \frac{\gamma_n}{2} Q b' \cos^2 (\theta_{B'}-\psi).
\ee
The vdW interaction free energy per unit area is then given in the non-retardation limit by
\ba
G_{sm} &\!\!=\!\!& 
\frac{k_{\rm{B}}T}{4\pi^2} \! \sum_n^\prime \! 
\int_0^{2\pi} \!\!\!\! d\psi \!\int_0^\infty \!\!\!\!dQ \, Q \, 
\ln \frac{\Theta_{11}^{(\rm{sm})}(d)}{\Theta_{11}^{(\rm{sm})}(d\rightarrow\infty)}
\nonumber\\
&\!\!=\!\!& 
\frac{k_{\rm{B}}T}{4\pi^2} \! \sum_n^\prime \! 
\int_0^{2\pi} \!\!\!\! d\psi \!\int_0^\infty \!\!\!\!dQ \, Q \,
\ln(1 - \bar{\Delta}_{W B_1} \bar{\Delta}^{\rm{(eff)}} e^{-2Qd}).
\nonumber\\
\label{eq:Gsm1}
\ea
Here we have defined an effective dielectric reflection coefficient to characterize the alternating $B'$- and $B$-type layers to the right of the intervening medium $m$:
\ba
\bar{\Delta}^{\rm{(eff)}} 
&\equiv& 
\frac{s_0 + s_1 \bar{\Delta}_{WB'} + s_2 \bar{\Delta}_{WB'}^2}{t_0 + t_1 \bar{\Delta}_{WB'} + t_2 \bar{\Delta}_{WB'}^2} 
\nonumber\\
&&\times 
\frac{2e^{-Q b' g_{B_1}} \sinh(Q b' g_{B_1})}{1 - \bar{\Delta}_{W B_1}^2 e^{-2Q b' g_{B_1}}},
\label{eq:Deff_exact}
\ea
where coefficients in the numerator are given in Eqs.~(\ref{eq:scoeffs}) and coefficients in the denominator are given in Eqs.~(\ref{eq:tcoeffs}) in App.~\ref{app:1}. 

The formulas Eqs.~(\ref{eq:Gsm1}) and (\ref{eq:Deff_exact}) are exact, which can be used to determine the interaction free energy behavior for arbitrary anisotropy strengths $\gamma_n$ and number of layers $N$. 
Equation~(\ref{eq:Gsm1}) is also formally equivalent to a vdW free energy of two interacting planar slabs, in which the effect of the \emph{multi-layeredness} of the second slab \emph{only} enters through the effective reflection coefficient, $\bar{\Delta}^{{(\rm eff)}}$. The logarithmic form of the free energy implies that it accounts for microscopic many-body effects to all orders. An expansion of the logarithm to quadratic order in reflection coefficients would correspond to making a Hamaker pairwise-summation approximation.

\subsection{van der Waals interaction free energy}

\begin{figure}
		\includegraphics[width=0.46\textwidth]{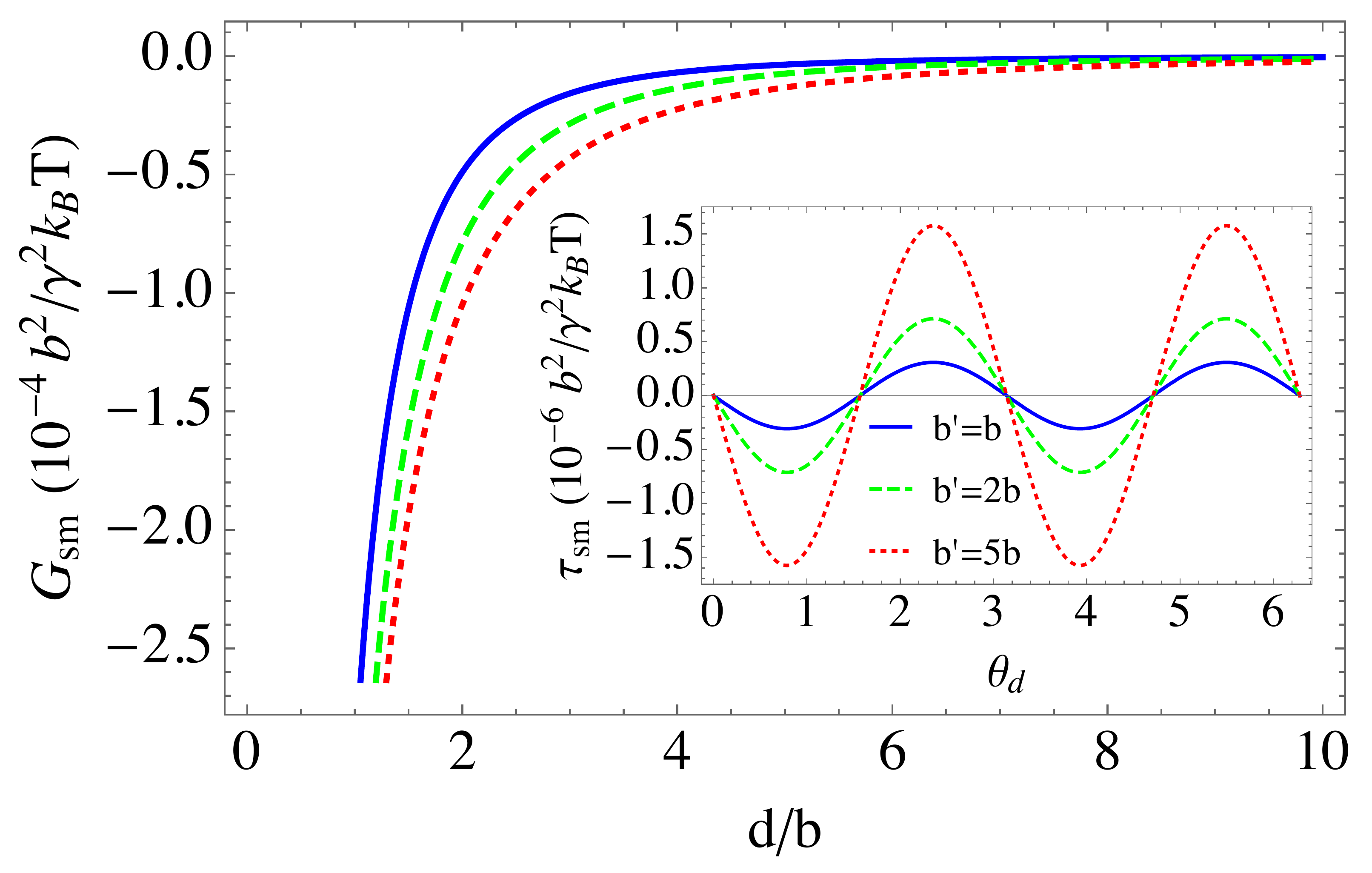}
	\caption{A single anisotropic layer interacting with a multilayer (weak anisotropy and large $N$): behavior of free energy per unit area $G_{sm}$ (Eq.~(\ref{eq:Gsm2})) with separation $d$ for $\theta_d=0$, and (inset) behavior of torque per unit area $\tau_{sm}$ (Eq.~(\ref{eq:tsm})) with $\theta_d$ for $d=10b$. Both behaviors are plotted for the following values of $b'$: (i)~$b'=b$ (blue), (ii)~$b'=2b$ (green, dashed), and (iii)~$b'=5b$ (red, dotted).}
\label{fig:sm}
\end{figure}

For $N \gg 1$, we have $U_{N-1} \approx e^{(N-1) \xi}$, and thus $w \approx e^{-\xi}$. 
For the case of weak anisotropy ($\gamma_n \ll 1$) we have $\bar{\Delta}_{W B_1} \approx -\gamma_n \cos^2(\theta_d-\psi)/4$ and $\bar{\Delta}_{WB'} \approx -\gamma_n (\cos\psi)^2/4$, i.e., $\bar{\Delta}_{WB'}$ is of the order of $\gamma_n$ and we can thus expand $\bar{\Delta}^{(\rm{eff})}$ in powers of $\gamma_n$. To leading order we find 
\be
\label{eq:Deff_approx}
\bar{\Delta}^{(\rm{eff})} 
\approx 
-\frac{\gamma_n \cos^2 \psi \, e^{-Q(b'-b)} \sinh^2(Q b')}{2\sinh(Q(b+b'))}. 
\ee
From Eq.~(\ref{eq:Gsm1}) we then find that the interaction free energy per unit area is given to the same order by 
\ba
\label{eq:Gsm2}
G_{sm} &\!\!\approx\!\!& 
-\frac{k_{{\rm B}}T}{4\pi^2} \! \sum_n^\prime \!
\int_0^{2\pi} \!\!\!\! d\psi \!\int_0^\infty \!\!\!\!dQ \, Q \, 
\bar{\Delta}_{W B_1} \bar{\Delta}^{({\rm eff})} 
e^{-2Q d}
\nonumber\\
&\!\!=\!\!&
-\frac{\gamma^2 k_{\rm{B}}T (1+2(\cos \theta_d)^2)}{2048\pi (b+b')^2}
\bigg[ 
\psi^{(1)}\left( \frac{d}{b+b'} \right) 
\nonumber\\
&&\quad- 
2 \psi^{(1)}\left( \frac{d+b'}{b+b'} \right) + \psi^{(1)}\left( \frac{d+2b'}{b+b'} \right)
\bigg]
\ea
where $\gamma^2 \equiv 2\sum_n^\prime \gamma_n^2$, $\psi^{(1)}(z) \equiv \partial \psi(z) / \partial z$ is the polygamma function of order unity, and $\psi(z) \equiv \Gamma'(z)/\Gamma(z)$ is the digamma function. 
We plot the behavior of $G_{sm}$ in Fig.~\ref{fig:sm}. The interaction is less attractive if the thickness $b$ of the $B$-type layers is larger, but becomes more attractive if the thickness $b'$ of the $B'$-type layers is larger. We can understand this as a manifestation of the vdW attraction being generated by the dielectric contrast between adjacent media. We have assumed that the dielectric permittivity of the $B$-type medium is identical to two of the principal permittivities of the anisotropic, $B'$-type medium, whilst the latter medium has an extra principal permittivity with a different value. Thus increasing $b$ ($b'$) implies a weakening (strengthening) of the effect of the dielectric contrast, and therefore the vdW attraction. 

In the large separation limit ($d \gg b+b'$) we find
\be
G_{sm}
\approx
-\frac{\gamma^2 k_{\rm{B}}T (1+2\cos^2 \theta_d) (b')^2}{1024\pi (b+b') d^3}, 
\ee
which corresponds to a force per unit area that decays with $d^{-4}$ and is given by 
\be
\mathcal{F}_{sm}
\approx
-\frac{3\gamma^2 k_{\rm{B}}T (1+2\cos^2 \theta_d) (b')^2}{1024\pi (b+b') d^4}. 
\ee
On the other hand, at small separations ($d \ll b+b'$) we find 
\be
G_{sm}
\approx 
-\frac{\gamma^2 k_{\rm{B}}T (1+2\cos^2 \theta_d)}{2048\pi d^2}.
\label{eq:Gsm_near}
\ee
This corresponds to a force per unit area which decays with $d^{-3}$ and is given by 
\be
\mathcal{F}_{sm}
\approx 
-\frac{\gamma^2 k_{\rm{B}}T (1+2\cos^2 \theta_d)}{1024\pi d^3}.
\ee
As expected Eq.~(\ref{eq:Gsm_near}) agrees with Eq.~(\ref{eq:Gss_near}), which effectively approximates the interaction behavior of two very thick anisotropic $B'$-type slabs. In contrast to the case of interacting single-layer $B'$-type plates, the behavior of $G_{sm}$ crosses over from $d^{-2}$ decay at small separation to $d^{-3}$ decay at large separation. 
Thus, for a single-layer $B'$-type plate interacting with a multilayered slab, the attraction at large separation is stronger than that for a pair of interacting single-layer $B'$-type plates. 

\subsection{van der Waals torque}

The van der Waals torque per unit area is given by 
\be
\tau_{sm} = -\frac{\partial G_{sm}}{\partial \theta_d}
\ee
In the weak anisotropy regime and for large $N$, we can compute $\tau_{sm}$ from Eq.~(\ref{eq:Gsm2}), obtaining 
\ba
\tau_{sm} &\!\!\approx\!\!& -\frac{\gamma^2 k_{{\rm B}}T \sin(2\theta_d)}{1024\pi (b+b')^2} 
\bigg[ 
\psi^{(1)}\left( \frac{d}{b+b'} \right) 
\nonumber\\
&&\quad- 
2 \psi^{(1)}\left( \frac{d+b'}{b+b'} \right) + \psi^{(1)}\left( \frac{d+2b'}{b+b'} \right)
\bigg]
\label{eq:tsm}
\ea
For large separation ($d \gg b+b'$), the torque is approximately given by 
\be
\tau_{sm} \approx -\frac{\gamma^2 k_{{\rm B}}T (b')^2 \sin(2\theta_d)}{512\pi (b+b') d^3}. 
\ee
For small separation ($d \ll b+b'$), we find
\be
\tau_{sm} \approx -\frac{\gamma^2 k_{{\rm B}}T \sin(2\theta_d)}{1024\pi d^2}. 
\ee
In Fig.~\ref{fig:sm}, we plot the behavior of the vdW torque as a function of $\theta_d$, for three different thicknesses $b'$. We see that the torque becomes enhanced as $b'$ increases. This enhancement can be understood as originating from the greater extent of dielectric contrast that we have discussed above. 

\section{Single anisotropic layer interacting with a multilayer having rotating optic axes} 
\label{sec:sr}

\begin{figure}
		\includegraphics[width=0.48\textwidth]{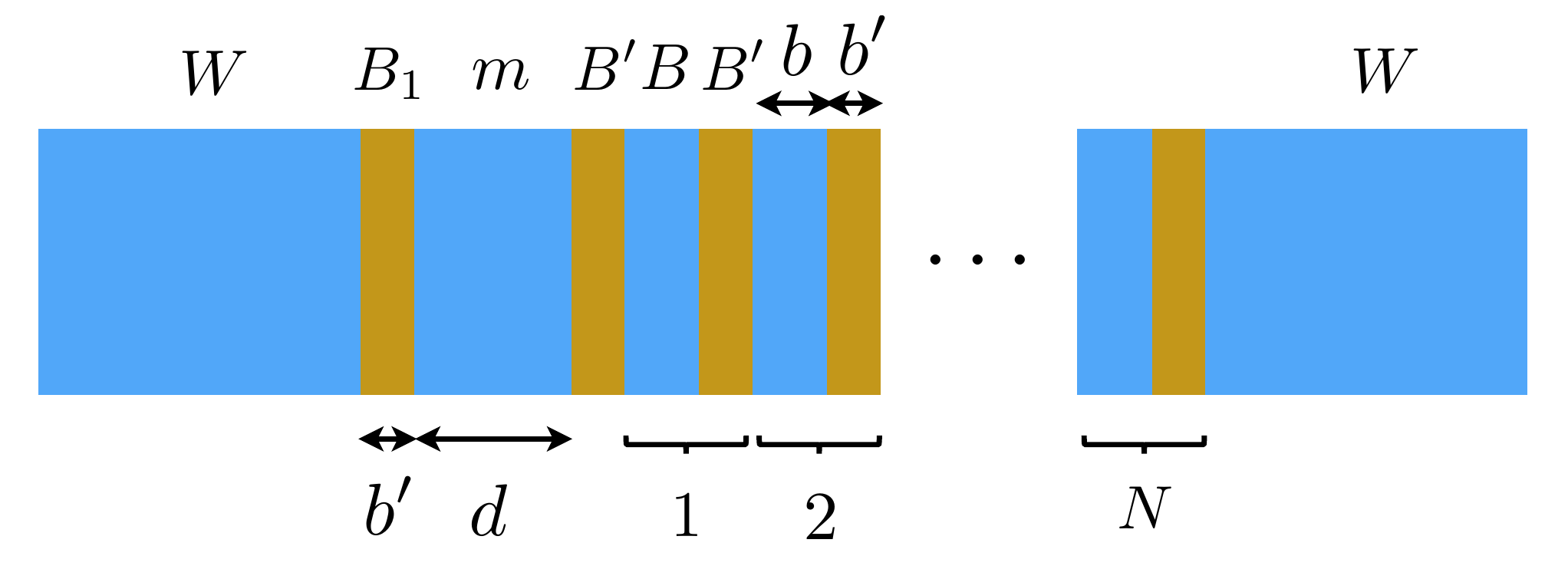}
	\caption{A single anisotropic layer $B_1$, oriented at angle $\theta_d$, is of thickness $b'$ and interacts with a slab composed of a sequence of alternating $B'$-type and $B$-type layers, of thicknesses $b'$ and $b$ respectively, across an intervening isotropic medium $m$ of thickness $d$. The layer $B_1$ and the slab are bounded on the left and the right respectively by isotropic media $W$. The numbers $1$, $2, \ldots, N$ refer to the $BB'$ bilayers, with optic axes oriented at $\delta\theta$, $2\delta\theta, \ldots N\delta\theta$ respectively. The optic axis of the $B'$-type layer immediately to the left of bilayer $1$ are oriented at zero angle.}
\label{fig:system_sr}
\end{figure}
Having considered a single anisotropic layer interacting with a multilayer in which the optic axes of all its anisotropic layers are aligned, we now turn to the case where the optic axes of the multilayer undergo angular increments of $\delta\theta$ on moving across the anisotropic, $B'$-type layers (see Figs.~\ref{fig:anisotropy} and \ref{fig:system_sr}). In the language of transfer matrices, this system is described by
\be
\begin{pmatrix}
\bing{\widetilde{A}_{R}} \\ \bing{\widetilde{B}_{R}}
\end{pmatrix} = 
\mathbf{\Theta}^{(\rm{sr})} \cdot
\begin{pmatrix}
\bing{\widetilde{A}_{L}} \\ 0
\end{pmatrix},
\ee
where
\ba
\mathbf{\Theta}^{(\rm{sr})} &\equiv& 
\left\{ \prod_{j=1}^{N} \mathbf{A}^{(j)} \right\}
\mathbf{D}_{WB'}^{(0)} \mathbf{T}_{B'}^{(0)} \mathbf{D}_{B'W}^{(0)} \mathbf{T}_{m} \mathbf{D}_{W B_1} 
 \mathbf{T}_{B_1} \mathbf{D}_{B_1 W}, 
\nonumber\\
\label{eq:Theta_sr}
\\
\mathbf{A}^{(j)} &\equiv& \mathbf{D}_{WB'}^{(j)} \mathbf{T}_{B'}^{(j)} \mathbf{D}_{B'W}^{(j)} \mathbf{T}_{B}, 
\label{eq:Ai_matrix}
\\
\mathbf{D}_{WB'}^{(j)} 
&\equiv&
\begin{pmatrix}
1 & -\bar{\Delta}_{WB'}^{(j)} \\
-\bar{\Delta}_{WB'}^{(j)} & 1
\end{pmatrix},
\label{eq:DWB'j}
\\
\label{eq:DeltaWBpj}
\bar{\Delta}_{WB'}^{(j)} &\equiv& \frac{1 - g_{B'}^{(j)}}{1 + g_{B'}^{(j)}}, 
\\
\label{eq:gBpj}
g_{B'}^{(j)} &\equiv& \sqrt{1 + \gamma_n \cos^2(j\delta\theta - \psi)} 
\\
\mathbf{T}_{B'}^{(j)} 
&\equiv& 
\begin{pmatrix}
1 & 0 \\
0 & e^{-2 Q b' g_{B'}^{(j)}}
\end{pmatrix}, 
\label{eq:TB'j}
\ea
As before, $\gamma_n \equiv \varepsilon_{B'x}(i\xi_n)/\varepsilon_{B'y}(i\xi_n) - 1$. The matrix product is ordered in the following manner
\be
\label{eq:prodAorder}
\prod_{j=1}^{N} \mathbf{A}^{(j)} \equiv \mathbf{A}^{(N)} \cdot \mathbf{A}^{(N-1)} \cdots \mathbf{A}^{(1)},
\ee 
and $\mathbf{D}_{W B_1}$, $\mathbf{T}_{B_1}$, $\mathbf{T}_m$ are defined by Eqs.~(\ref{eq:DWL}), (\ref{eq:TL}) and (\ref{eq:Tm}). We set the orientation of the optic axis of the $B'$-type layer immediately to the right of medium $m$ to be at zero angle, so $\theta_d$ is the relative orientation of this axis with respect to the axis of slab $B_1$. The values of the matrix elements of $\mathbf{A}^{(j)}$ are given in Eqs.~(\ref{eq:Asla}) of App.~\ref{app:2}. 
As in the case of interacting single-layered slabs, the vdW interaction free energy per unit area is given by 
\be
G = \frac{k_{\rm{B}}T}{4\pi^2} \! \sum_n^\prime \! \int \! dQ\,Q \int \! d\psi \,\ln \frac{\Theta_{11}^{(\rm{sr})}(d,i\xi_n)}{\Theta_{11}^{(\rm{sr})}(d\rightarrow\infty,i\xi_n)}
\ee
The matrix $\Theta^{({{\rm sr}})}$ involves a product over $N$ matrices $\mathbf{A}^{(j)}$, each of which depends on the specific orientation of the optic axis of the layer. 

For a slab with a large number of layers, it is probably not possible to obtain an exact closed-form result for the free energy per unit area. We also do not have the benefit of Abel\`{e}s' formula which is valid only for the product of $N$ identical matrices. However it is still possible to obtain a relatively simple expression for the case of weak anisotropy ($\gamma \ll 1$), where the expression involves a single momentum integral. In the case where $\delta\theta=0$, we will obtain a free energy expression completely in terms of analytic functions, and equivalent to the free energy expression we already obtained in Sec.~\ref{sec:sm}. 

In what follows, we consider the weak anisotropy regime. 
In this regime, we can approximate Eqs.~(\ref{eq:DWL}), (\ref{eq:gL}), (\ref{eq:DeltaWBpj}) and (\ref{eq:gBpj}) by
\ba
\bar{\Delta}_{W B_1} &\approx& -\frac{\gamma_n}{4} (\cos(\theta_d-\psi))^2
\label{eq:DeltaWLappr}
\\
g_{B_1} &\approx& 1 + \frac{\gamma_n}{2} (\cos(\theta_d-\psi))^2
\\
\bar{\Delta}_{WB'}^{(j)} &\approx& -\frac{\gamma_n}{4} (\cos(j\delta\theta - \psi))^2
\\
g_{B'}^{(j)} &\approx& 1 + \frac{\gamma_n}{2} (\cos(j\delta\theta-\psi))^2
\ea
Similarly, we can approximate 
\be
\label{eq:Ajappr}
\mathbf{A}^{(j)} \approx \mathbf{A}_0 + \delta\mathbf{A}^{(j)},
\ee
where $\mathbf{A}_0$ and $\delta\mathbf{A}^{(j)}$ are of zeroth and linear order in $\gamma_n$ respectively, and we can make a corresponding linear-order approximation to the matrix product 
\be
\label{eq:B}
\prod_{j=1}^{N} \mathbf{A}^{(j)} \approx \mathbf{A}_0^N + \mathbf{B}, 
\ee
where $\mathbf{B}$ is a matrix of linear order in $\gamma$, and encodes the effect of the dielectric anisotropy:
\ba
\mathbf{B} &\equiv& \sum_{j=1}^{N} \mathbf{A}_0^{N-j} \delta\mathbf{A}^{(j)}\mathbf{A}_0^{j-1}
\\
&=&
\delta\mathbf{A}^{(N)}\mathbf{A}_0^{N-1} + \mathbf{A}_0\delta\mathbf{A}^{(N-1)}\mathbf{A}_0^{N-2} 
\nonumber\\
&&+ \dots + \mathbf{A}_0^{N-2}\delta\mathbf{A}^{(2)}\mathbf{A}_0 + \mathbf{A}_0^{N-1}\delta\mathbf{A}^{(1)}
\nonumber
\ea
The matrix elements of $\mathbf{A}_0$, $\delta\mathbf{A}^{(j)}$ and $\mathbf{B}$ can be computed and are given by Eqs.~(\ref{eq:Amatrixelements}), (\ref{eq:deltaAele}) and (\ref{eq:Bele}) of App.~\ref{app:2}. 
We can rewrite $\mathbf{\Theta}^{{\rm rr}}$ in Eq.~(\ref{eq:Theta_sr}) in the following form: 
\be
\mathbf{\Theta}^{(\rm{sr})} = \mathbf{\Theta}^{(R)} \cdot \mathbf{T}_{m} \cdot \mathbf{\Theta}^{(L)}, 
\ee
where 
\begin{subequations}
\label{eq:Thetasinglemany}
\ba
\mathbf{\Theta}^{(R)} &\equiv& \left\{ \prod_{j=1}^{N} \mathbf{A}^{(j)} \right\}
\mathbf{D}_{WB'}^{(0)} \mathbf{T}_{B'}^{(0)} \mathbf{D}_{B'W}^{(0)}; 
\\
\mathbf{\Theta}^{(L)} &\equiv& 
\mathbf{D}_{W B_1} \mathbf{T}_{B_1} \mathbf{D}_{B_1 W}
\ea
\end{subequations}
The matrix elements of $\mathbf{\Theta}^{(R)}$ and $\mathbf{\Theta}^{(L)}$ are given in Eqs.~(\ref{eq:A1}) in App.~\ref{app:2}. 
The matrix element $\Theta_{11}^{({{\rm sr}})}$ is given by
\be
\Theta_{11}^{({{\rm sr}})} = \Theta_{11}^{(L)} \Theta_{11}^{(R)} + \Theta_{21}^{(L)} \Theta_{12}^{(R)} e^{-2Q d}
\ee
and the normalized dispersion relation is given by 
\be
\frac{\Theta_{11}^{({{\rm sr}})}(d,i\xi_n)}{\Theta_{11}^{({{\rm sr}})}(d\rightarrow\infty,i\xi_n)} 
= 1 - \bar{\Delta}_{WR}^{{\rm (eff)}}(i\xi_n) \bar{\Delta}_{W B_1}^{{\rm (eff)}}(i\xi_n) e^{-2Q d},
\ee
where the effective reflection coefficients corresponding to the interfaces of the left and of the right slabs with the intervening \bing{solvent} medium are given by 
\be
\bar{\Delta}_{WR}^{{\rm (eff)}} \equiv \frac{\Theta_{12}^{(R)}}{\Theta_{11}^{(R)}}, \quad 
\bar{\Delta}_{W B_1}^{{\rm (eff)}} \equiv - \frac{\Theta_{21}^{(L)}}{\Theta_{11}^{(L)}}.
\ee
In the weak anisotropy limit, these coefficients can be further approximated by 
\begin{subequations}
\label{eq:fresnels}
\ba
\bar{\Delta}_{WR}^{{\rm (eff)}} 
&\approx& 
-\frac{\gamma_n}{2}e^{-Q b'}\sinh(Q b')\cos^2\psi 
\nonumber\\
&&- \frac{\gamma_n e^{-Q b'} \sinh(Q b')}{8(\cosh(2Q(b+b'))-\cos(2\delta\theta))}
\nonumber\\
&&\quad\times 
\big[ 
\cos(2(\delta\theta-\psi)) - e^{-2Q(b+b')}\cos(2\psi)
\nonumber\\
&&\qquad+ 
e^{-2(N+1)Q(b+b')} \cos(2(N\delta\theta-\psi))
\nonumber\\
&&\qquad-
e^{-2NQ(b+b')} \cos(2((N+1)\delta\theta-\psi))
\big]
\nonumber\\
&&-\frac{\gamma_n}{4} e^{-Q((N+1)b+(N+2)b')} \sinh(Q b') 
\nonumber\\
&&\quad\times
\frac{\sinh(NQ(b+b'))}{\sinh(Q(b+b'))}
\\
\bar{\Delta}_{W B_1}^{{\rm (eff)}} 
&\approx& 
-\frac{\gamma_n}{2}e^{-Q b'} \sinh(Q b') \cos^2(\theta_d-\psi).
\ea
\end{subequations}

\subsection{van der Waals free energy}

The interaction free energy per unit area is then given by 
\ba
G_{sr} 
&\!\!=\!\!&
\frac{k_{\rm{B}}T}{4\pi^2} \! \sum_n^\prime \! 
\int_0^{2\pi} \!\!\!\! d\psi \!\int_0^\infty \!\!\!\!dQ \, Q \, 
\ln \frac{\Theta_{11}^{(\rm{sr})}(d,i\xi_n)}{\Theta_{11}^{(\rm{sr})}(d\rightarrow\infty,i\xi_n)}
\nonumber\\
&\!\!=\!\!& 
\frac{k_{\rm{B}}T}{4\pi^2} \! \sum_n^\prime \! \int \! dQ\,Q \int \! d\psi \,\ln (1 - \bar{\Delta}_{W B_1}^{{\rm (eff)}} \bar{\Delta}_{WR}^{(\rm{eff})} e^{-2Qd})
\nonumber\\
&\!\!=\!\!&
\frac{k_{\rm{B}}T}{4\pi^2} \! \sum_n^\prime \! \int \! dQ\,Q \int \! d\psi \,\ln (1 - \bar{\Delta}_{W B_1} \bar{\Delta}^{(\rm{eff})} e^{-2Qd})
\nonumber\\
\label{eq:Gsr1}
\ea
Using Eqs.~(\ref{eq:fresnels}) and (\ref{eq:DeltaWLappr}) we find for $\bar{\Delta}^{{\rm (eff)}}$
\ba
\label{eq:Deff_n}
\bar{\Delta}^{(\rm{eff})} &\!\!=\!\!& 
-\gamma_n e^{-2Qb'} (\sinh Qb')^2 (\cos\psi)^2
\nonumber\\
&&-
\frac{\gamma_n (\sinh Qb')^2}{4(\cosh(2Q(b+b')) - \cos 2\delta\theta)}
\nonumber\\
&&\quad\times
\big[ 
e^{-2Q b'} \cos(2\delta\theta-2\psi) - e^{-2Q(b+2b')} \cos 2\psi
\nonumber\\
&&\qquad+
e^{-2Q((N+1)b+(N+2)b'))} \cos(2N\delta\theta-2\psi)
\nonumber\\
&&\qquad-
e^{-2Q(Nb+(N+1)b')} \cos(2(N+1)\delta\theta-2\psi)
 \big]
\nonumber\\
&&-
\frac{\gamma_n}{2} e^{-Q((N+1)b+(N+3)b')} 
\nonumber\\
&&\quad\times 
\frac{(\sinh Qb')^2 \sinh(NQ(b+b'))}{\sinh(Q(b+b'))}
\ea
In the large $N$ limit, the above simplifies to
\ba
&&\bar{\Delta}^{(\rm{eff})} 
\\
&\!\!\approx\!\!& 
-\gamma_n e^{-2Qb'} (\sinh Qb')^2 (\cos\psi)^2
\nonumber\\
&&-
\frac{\gamma_n (\sinh Qb')^2}{4(\cosh(2Q(b+b')) - \cos 2\delta\theta)}
\nonumber\\
&&\quad\times
\big[ 
e^{-2Q b'} \cos(2\delta\theta-2\psi) - e^{-2Q(b+2b')} \cos 2\psi
 \big]
\nonumber\\
&&-
\frac{\gamma_n e^{-Q(b+3b')} (\sinh Qb')^2}{4 \sinh(Q(b+b'))}
\ea
For the case $\delta\theta=0$, the above expression for $\bar{\Delta}^{(\rm{eff})}$ reduces to Eq.~(\ref{eq:Deff_approx}), as we expect. 
\begin{figure}
		\includegraphics[width=0.48\textwidth]{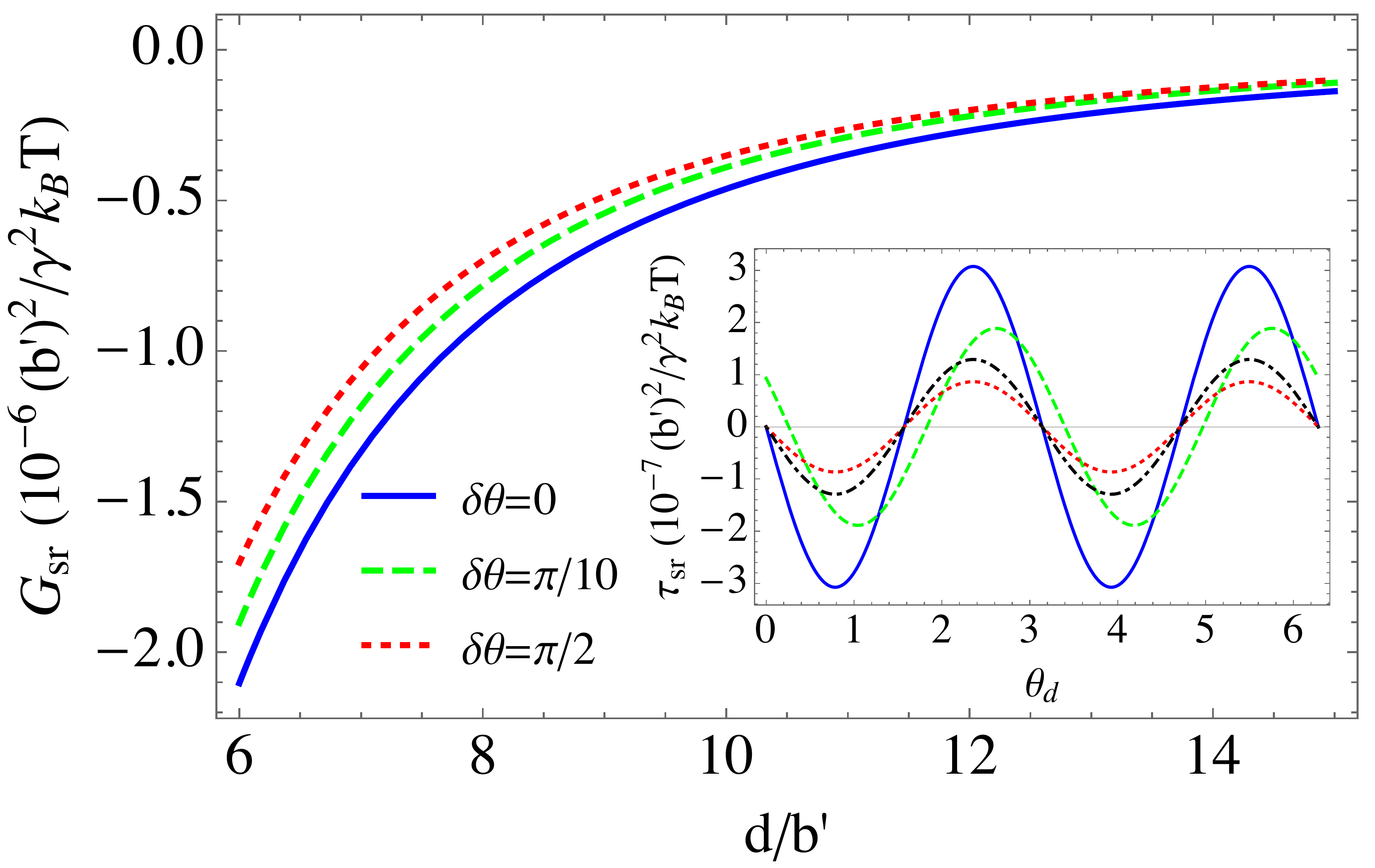}
\caption{Single anisotropic layer interacting with multilayer having rotating optic axes (weak anisotropy, large $N$, and $b'=b$): behavior of $G_{sr}$ (Eq.~(\ref{eq:Gsr})) with $d$ for $\theta_d=0$, and (inset) behavior of $\tau_{sr}$ (Eq.~(\ref{eq:tsr})) with $\theta_d$ for $d=10b'$, for the following values of $\delta\theta$: (i)~$\delta\theta=0$ (blue), (ii)~$\delta\theta=\pi/10$ (green, dashed), and (iii)~$\delta\theta=\pi/2$ (red, dotted). For comparison we have shown $\tau_{ss}$ for $\theta_d=0$ (black, dot-dashed line; cf. Eq.~(\ref{eq:tss})).}	
\label{fig:manybody}
\end{figure}
For weak anisotropy we can also approximate the free energy per unit area to leading order by
\be
G_{sr} \approx -\frac{k_{\rm{B}}T}{4\pi^2} \! \sum_n^\prime \! \int \! dQ\,Q \int \! d\psi \, \bar{\Delta}_{W B_1} \bar{\Delta}^{(\rm{eff})} e^{-2Qd}
\ee
In the large $N$ limit, we find 
\be
G_{sr} \approx G_{ss} + \delta G_{sr}
\label{eq:Gsr}
\ee
where $G_{ss}$ is the interaction free energy per unit area of two thin single-layered slabs (cf. Sec.~\ref{sec:singlelayers}), given by 
\ba
G_{ss} &\equiv&
-\frac{\gamma^2 k_{{\rm B}}T (1+2(\cos\theta_d)^2)}{2048\pi} 
\nonumber\\
&&\times 
\left[ \frac{1}{d^2} - \frac{2}{(d+b')^2} + \frac{1}{(d+2b')^2} \right], 
\label{eq:Gsr0}
\ea
and $\delta G_{sr}$ is the correction from the additional layers of the second slab, given by
\ba
&&\delta G_{sr} 
\equiv -
\frac{\gamma^2 k_{{\rm B}}T}{1024\pi(b+b')^2} 
\bigg[
\psi^{(1)}\left(\frac{d+b+b'}{b+b'} \right)
\nonumber\\
&&\quad-
2\psi^{(1)}\left(\frac{d+b+2b'}{b+b'} \right)
+\psi^{(1)}\left(\frac{d+b+3b'}{b+b'} \right)
\bigg]
\nonumber\\
&&-
\frac{\gamma^2 k_{{\rm B}}T}{256\pi} 
\!  \int \! dQ\, Q
\bigg[
\frac{e^{-2Q b'}(\sinh Q b')^2 \cos(2(\delta\theta - \theta_d))}{\cosh(2Q(b+b')) - \cos 2\delta\theta}
\nonumber\\
&&\quad-
\frac{e^{-2Q(b+2b')}(\sinh Q b')^2 \cos(2 \theta_d)}{\cosh(2Q(b+b')) - \cos 2\delta\theta}
\bigg] e^{-2 Q d}
\label{eq:dGsr}
\ea
For the case where the optic axis of each layer in the second slab are oriented at the same angle (i.e., $\delta\theta=0$), the interaction free energy per unit area admits of a closed-form expression: 
\ba
G_{sr} &\!\!\approx\!\!& 
-\frac{\gamma^2 k_{{\rm B}}T (1+2(\cos\theta_d)^2)}{2048\pi} 
\nonumber\\
&&\quad\times 
\left[ \frac{1}{d^2} - \frac{2}{(d+b')^2} + \frac{1}{(d+2b')^2} \right]
\nonumber\\
&&-\frac{\gamma^2 k_{{\rm B}}T (1+2(\cos\theta_d)^2)}{2048\pi(b+b')^2} 
\bigg[
\psi^{(1)}\left( \frac{d+b+b'}{b+b'} \right)
\nonumber\\
&&\quad-
2\psi^{(1)}\left( \frac{d+b+2b'}{b+b'} \right)
+\psi^{(1)}\left( \frac{d+b+3b'}{b+b'} \right)
\bigg]
\nonumber\\
\ea
As expected, this is in fact equivalent to Eq.~(\ref{eq:Gsm2}), i.e., the interaction free energy per unit area of a single anisotropic layer interacting with a multilayer having optic axes all aligned. 
On the other hand, as we progressively increase the thickness of the isotropic, $B$-type layers within the multilayered slab (i.e., let $b \rightarrow \infty$), we expect to recover the interaction energy of two thin single-layered slabs; this is indeed the case as we see from Eq.~(\ref{eq:dGsr}), in which the correction $\delta G_{sr} \rightarrow 0$, and thus $G_{sr} \rightarrow G_{ss}$. 

\subsection{van der Waals torque}
\label{sec:vdwtorquesr}

Using Eqs.~(\ref{eq:Gsr}), (\ref{eq:Gsr0}) and (\ref{eq:dGsr}), we obtain the torque per unit area $\tau_{sm}$ for a single layered slab interacting with a multi-layered slab:
\be
\tau_{sr} = \tau_{ss} + \delta\tau_{sr}, 
\label{eq:tsr}
\ee
where 
\ba
\delta\tau_{sr} &\equiv& 
- \frac{\gamma^2 k_{{\rm B}}T}{128\pi} 
\! \int \! dQ \, Q \, e^{-2Q b' - 2Q d} 
\\
&&\quad\times 
\frac{(\sinh Qb')^2}{\cos 2\delta\theta - \cosh (2Q(b+b'))} 
\nonumber\\
&&\quad\times 
[
\sin(2 (\delta\theta - \theta_d)) 
+ 
e^{-2Q(b+b')} \sin(2\theta_d)
].
\nonumber
\ea
In Fig.~\ref{fig:manybody}, we consider the effect of changing $\delta\theta$ on the vdW torque (inset) and the interaction free energy for the case where $\theta_d=0$. The vdW attraction is strongest and the torque has the maximum amplitude for $\delta\theta=0$, progressively becoming weaker as $\delta\theta$ increases to $\pi/2$. We see that whilst the vdW torque for $\delta\theta = 0$ (shown as the blue curve) is stronger than that for two interacting single anisotropic layers (shown as the black dot-dashed curve), as we increase $\delta\theta$ the vdW torque weakens and can in fact become smaller than that for the two single layers, as we see from the behavior for $\delta\theta=\pi/2$ (shown as the red dotted curve).

We may intuitively understand this in the following manner. For $\delta\theta=0$, every anisotropic layer in the multilayer will experience the same deviation of $\theta_d$ and hence a torque with the same sign, so the overall torque that acts on the multilayer is enhanced relative to that acting on a single anisotropic layer. Conversely, for $\delta\theta=\pi/2$, a perturbation of $\theta_d$ will cause half the anisotropic layers in the multilayer to experience an attractive torque and the other half to experience a repulsive torque, so the overall torque acting on the multilayer as a whole will be smaller than that acting on a single layer. This overall torque is \emph{not} however equal to zero, because the \emph{magnitude} of the torque is different for layers at different positions, becoming smaller for the ones that are farther away. For $\delta\theta=0$ and $\delta\theta=\pi/2$, the stable (unstable) angular configurations are those for which $\theta_d=n\pi$ ($\theta_d=(n+\tfrac{1}{2})\pi$), where $n$ is integer. Here we define the stable (unstable) angular configuration as one for which the torque is zero, and the multilayer experiences an attractive (repulsive) torque when $\theta_d$ is perturbed. On the other hand, as we increase $\delta\theta$ from $0$ to $\pi/2$, the torque amplitude decreases and additionally there is a ``phase shift" as the angular positions of the stable and unstable configurations take on values different from $n\pi$ and $(n+\tfrac{1}{2})\pi$.  

\section{Two interacting multilayers with rotating optic axes}
\label{sec:rr}

\begin{figure}
		\includegraphics[width=0.49\textwidth]{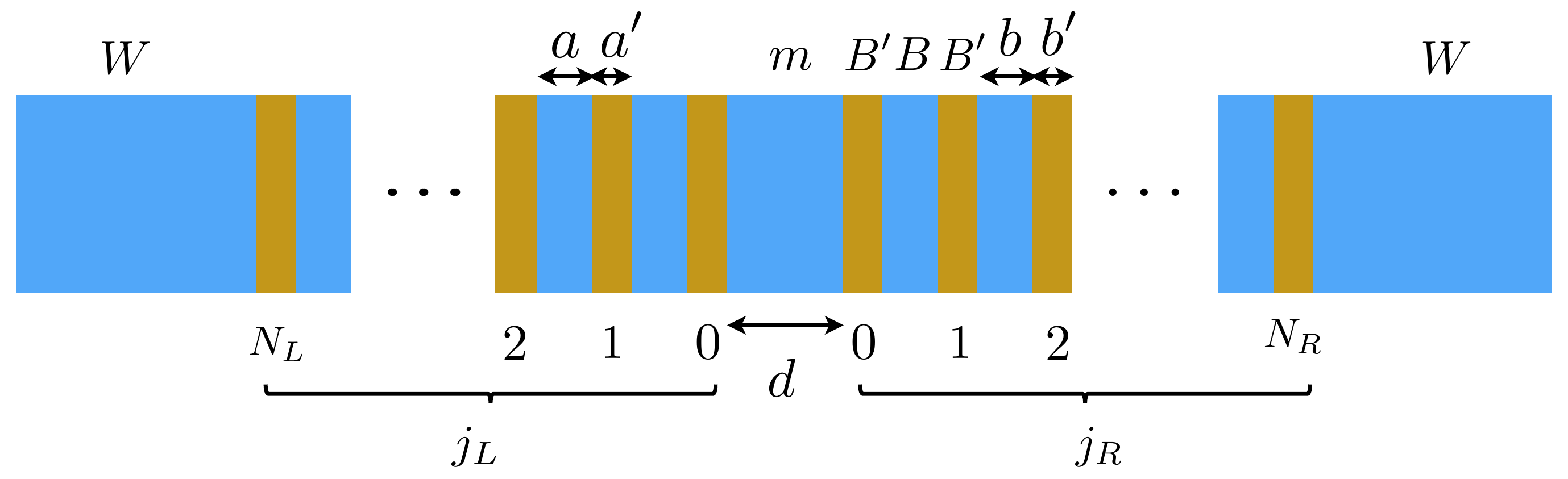}
	\caption{Two multilayered slabs, each consisting of alternating $B'$-type and $B$-type layers, interacting across an isotropic medium $m$ of thickness $d$.  The left (right) slab has $N_L+1$ ($N_R+1$) $B'$-type layers, and the \bing{solvent} (blue) and $B'$-type (brown) layers in the left (right) slab have thicknesses $a$ and $a'$ ($b$ and $b'$) respectively.  The label $j_L$ ($j_R$) is an index for $B'$-type layers in the left (right) slab. In the right slab, the optic axis of the layer at $j_R=0$ is oriented at zero angle and the the orientation of the optic axis of each successive layer on the right is $j_R\delta\theta$. In the left slab, the optic axis of the layer at $j_L=0$ is oriented at $\theta_d$ and the the orientation of the optic axis of each successive layer on the left is $\theta_d - j_L\delta\theta$. The two slabs thus have the same chirality, i.e., the optic axis rotates \emph{clockwise} as one moves from left to right.}
\label{fig:system_rr}
\end{figure}

We can straightforwardly generalize our results to the case of two multilayered slabs interacting across an isotropic medium. Let us consider a system consisting of two slabs separated by a \bing{solvent} medium of thickness $d$. The left (right) slab has $N_L + 1$ ($N_R + 1$) $B'$-type layers each of thickness $a'$ ($b'$) and $N_L$ ($N_R$) \bing{solvent} layers each of thickness $a$ ($b$). The optic axis of the $B'$-type layer in the left (right) slab immediately adjacent to the intervening medium is oriented at an angle $\theta_d$ ($0$). For simplicity, we assume that the dielectric properties of the $B'$-type layers in both slabs are the same, and represent the system in terms of transfer matrices, viz., 
\be
\begin{pmatrix}
\bing{\widetilde{A}_{R}} \\ \bing{\widetilde{B}_{R}}
\end{pmatrix} = 
\mathbf{\Theta}^{(\rm{rr})} \cdot
\begin{pmatrix}
\bing{\widetilde{A}_{L}} \\ 0
\end{pmatrix}.
\ee
Here
\ba
\mathbf{\Theta}^{(\rm{rr})} &\equiv& 
\left\{ \prod_{j=1}^{\bing{N_R}} \mathbf{A}^{(j)} \right\}
\mathbf{D}_{WB'}^{(0)} \mathbf{T}_{B'}^{(0)} \mathbf{D}_{B'W}^{(0)} 
\mathbf{T}_{m} 
\nonumber\\
&&\cdot
\widetilde{\mathbf{D}}_{WB'}^{(0)} \widetilde{\mathbf{T}}_{B'}^{(0)} \widetilde{\mathbf{D}}_{B'W}^{(0)} 
\left\{ \prod_{j=1}^{\bing{N_L}} \widetilde{\mathbf{A}}^{(j)} \right\}
\label{eq:Theta_rr}
\ea
with
\ba
\widetilde{\mathbf{A}}^{(j)} 
&\equiv& 
\mathbf{T}_{B} \widetilde{\mathbf{D}}_{WB'}^{(j)} \widetilde{\mathbf{T}}_{B'}^{(j)} 
\widetilde{\mathbf{D}}_{B'W}^{(j)}, 
\label{eq:tAj_matrix}
\\
\widetilde{\mathbf{D}}_{WB'}^{(j)} 
&\equiv&
\begin{pmatrix}
1 & -\widetilde{\Delta}_{WB'}^{(j)} \\
-\widetilde{\Delta}_{WB'}^{(j)} & 1
\end{pmatrix},
\\
\label{eq:tDeltaWBpj}
\widetilde{\Delta}_{WB'}^{(j)} &\equiv& \frac{1 - \widetilde{g}_{B'}^{(j)}}{1 + \widetilde{g}_{B'}^{(j)}}, 
\label{eq:tgBpj}
\ea
and
\ba
\widetilde{g}_{B'}^{(j)} &\equiv& \sqrt{1 + \gamma_n \cos^2(\theta_d-j\delta\theta - \psi)} 
\\
\widetilde{\mathbf{T}}_{B'}^{(j)} 
&\equiv& 
\begin{pmatrix}
1 & 0 \\
0 & e^{-2 Q a' \widetilde{g}_{B'}^{(j)}}
\end{pmatrix},
\\
\widetilde{\mathbf{T}}_{B} 
&\equiv& 
\begin{pmatrix}
1 & 0 \\
0 & e^{-2 Q a}
\end{pmatrix}. 
\ea
Above, $\mathbf{A}^{(j)}$, $\mathbf{D}_{WB'}^{(j)}$, $\bar{\Delta}_{WB'}^{(j)}$, $g_{B'}^{(j)}$ and $\mathbf{T}_{B'}^{(j)}$  are given by Eqs.~(\ref{eq:Ai_matrix}), (\ref{eq:DWB'j}), (\ref{eq:DeltaWBpj}), (\ref{eq:gBpj}) and (\ref{eq:TB'j}), respectively. The quantities $\mathbf{D}_{W B_1}$, $\mathbf{T}_{B_1}$, $\mathbf{T}_m$ are defined by Eqs.~(\ref{eq:DWL}), (\ref{eq:TL}) and (\ref{eq:Tm}). 

The matrix product $\prod_{j=1}^{\bing{N_R}} \mathbf{A}^{(j)}$ is ordered as in Eq.~(\ref{eq:prodAorder}), whereas $\prod_{j=1}^{\bing{N_L}} \widetilde{\mathbf{A}}^{(j)}$ is ordered in the contrary direction, viz., 
\be
\bing{\prod_{j=1}^{N_L} \widetilde{\mathbf{A}}^{(j)} \equiv \widetilde{\mathbf{A}}^{(1)} \cdot \widetilde{\mathbf{A}}^{(2)} \cdots \widetilde{\mathbf{A}}^{(N_L)}.}
\ee 
The matrix elements of $\widetilde{\mathbf{A}}^{(j)}$ are given in Eqs.~(\ref{eq:thales}) of App.~\ref{app:multilayers}. 
As in Sec.~\ref{sec:sr} we consider the weak anisotropy regime. 
We can thus approximate 
\ba
\widetilde{\Delta}_{WB'}^{(j)} &\approx& -\frac{\gamma_n}{4} (\cos(\theta_d-j\delta\theta - \psi))^2
\\
\widetilde{g}_{B'}^{(j)} &\approx& 1 + \frac{\gamma_n}{2} (\cos(\theta_d-j\delta\theta-\psi))^2.
\ea
Similarly, we can approximate 
\be
\label{eq:anne0}
\widetilde{\mathbf{A}}^{(j)} \approx \bing{\widetilde{\mathbf{A}}_0} + \delta\widetilde{\mathbf{A}}^{(j)},
\ee
where $\bing{\widetilde{\mathbf{A}}_0}$ and $\delta\widetilde{\mathbf{A}}^{(j)}$ are of zeroth and linear order in $\gamma_n$ respectively, and we can make a corresponding linear-order approximation to the matrix product 
\be
\prod_{j=1}^{\bing{N_L}} \widetilde{\mathbf{A}}^{(j)} \approx \bing{\widetilde{\mathbf{A}}_0^{N_L}} + \widetilde{\mathbf{B}}, 
\ee
where $\widetilde{\mathbf{B}}$ is a matrix of linear order in $\gamma_n$:
\ba
\label{eq:caroline0}
\widetilde{\mathbf{B}} 
&\equiv& 
\sum_{j=1}^{\bing{N_L}} \widetilde{\mathbf{A}}_0^{j-1} \delta\widetilde{\mathbf{A}}^{(j)} \widetilde{\mathbf{A}}_0^{\bing{N_L}-j}
\\
&=&
\delta\widetilde{\mathbf{A}}^{(1)}\mathbf{A}_0^{\bing{N_L}-1} 
+ \widetilde{\mathbf{A}}_0 \,
\delta\widetilde{\mathbf{A}}^{(2)} 
\widetilde{\mathbf{A}}_0^{\bing{N_L}-2}
\nonumber\\
&&+ \dots 
+ \widetilde{\mathbf{A}}_0^{\bing{N_L}-2} \delta\widetilde{\mathbf{A}}^{(\bing{N_L}-1)} 
\widetilde{\mathbf{A}}_0
+ \widetilde{\mathbf{A}}_0^{\bing{N_L}-1} \delta\widetilde{\mathbf{A}}^{(\bing{N_L})}
\nonumber
\ea
The matrix elements of \bing{$\widetilde{\mathbf{A}}_0$ are given by Eq.~(\ref{app:widetildeA0})}, whilst $\delta\widetilde{\mathbf{A}}^{(j)}$ and $\widetilde{\mathbf{B}}$ are given by Eqs.~(\ref{eq:anne}) and (\ref{eq:caroline}). 
We can rewrite $\mathbf{\Theta}^{{\rm rr}}$ in Eq.~(\ref{eq:Theta_rr}) as the following matrix product: 
\be
\mathbf{\Theta}^{(\rm{rr})} = \mathbf{\Theta}^{(R)} \cdot \mathbf{T}_{m} \cdot \mathbf{\Theta}^{(L)}, 
\ee
where 
\begin{subequations}
\label{eq:multilayerM}
\ba
\mathbf{\Theta}^{(R)} &\equiv& \left\{ \prod_{j=1}^{\bing{N_R}} \mathbf{A}^{(j)} \right\}
\mathbf{D}_{WB'}^{(0)} \mathbf{T}_{B'}^{(0)} \mathbf{D}_{B'W}^{(0)}; 
\\
\mathbf{\Theta}^{(L)} &\equiv& \widetilde{\mathbf{D}}_{WB'}^{(0)} \widetilde{\mathbf{T}}_{B'}^{(0)} \widetilde{\mathbf{D}}_{B'W}^{(0)} 
\left\{ \prod_{j=1}^{\bing{N_L}} \widetilde{\mathbf{A}}^{(j)} \right\}.
\ea
\end{subequations}
The matrix elements of $\mathbf{\Theta}^{(R)}$ and $\mathbf{\Theta}^{(L)}$ are given in Eqs.~(\ref{eq:multilayerelms}) in App.~\ref{app:multilayers}.  
\begin{figure}
		\includegraphics[width=0.48\textwidth]{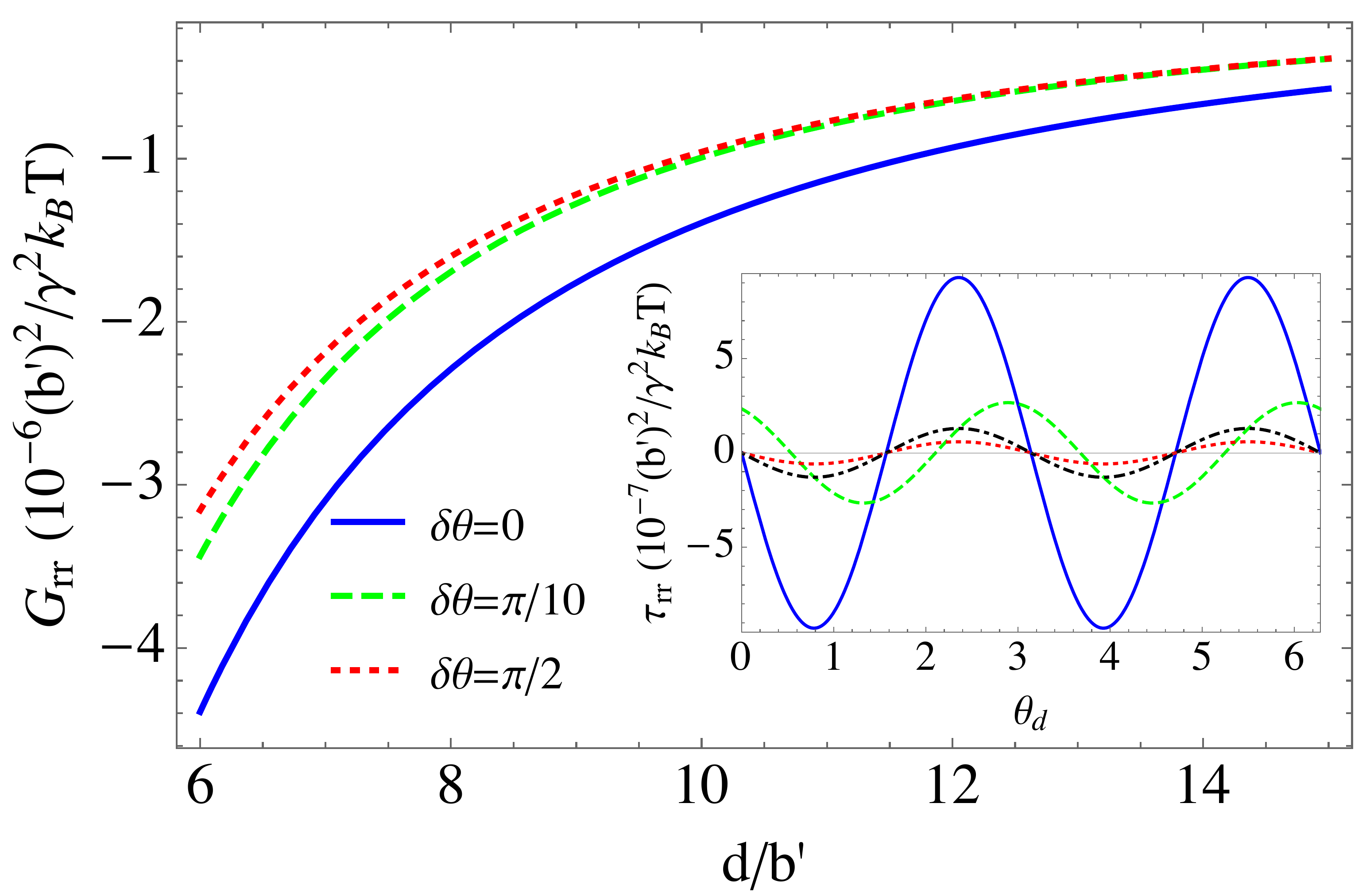}
	\caption{Two interacting multilayers with rotating optic axes (weak anisotropy, large $N$, and $a'=a=b'=b$): behavior of $G_{rr}$ (Eq.~(\ref{eq:Gmulti})) with $d$ for $\theta_d=0$, and (inset) behavior of $\tau_{rr}$ (Eq.~(\ref{eq:trr})) with $\theta_d$ for $d=10b'$, for the following values of $\delta\theta$: (i)~$\delta\theta=0$ (blue), (ii)~$\delta\theta=\pi/10$ (green, dashed), and (iii)~$\delta\theta=\pi/2$ (red, dotted). For comparison we have shown $\tau_{ss}$ for $\theta_d=0$ (black, dot-dashed line; cf. Eq.~(\ref{eq:tss})).}
\label{fig:rr}
\end{figure}
The matrix element $\Theta_{11}^{({{\rm rr}})}$ is given by
\be
\Theta_{11}^{({{\rm rr}})} = \Theta_{11}^{(L)} \Theta_{11}^{(R)} + \Theta_{21}^{(L)} \Theta_{12}^{(R)} e^{-2Q d}
\ee
and the dispersion relation is given by
\be
\frac{\Theta_{11}^{({{\rm rr}})}(d,i\xi_n)}{\Theta_{11}^{({{\rm rr}})}(d\rightarrow\infty,i\xi_n)} 
= 1 - \bar{\Delta}_{WR}^{{\rm (eff)}}(i\xi_n) \bar{\Delta}_{W B_1}^{{\rm (eff)}}(i\xi_n) e^{-2Q d}
\ee
where 
\be
\label{eq:yael}
\bar{\Delta}_{WR}^{{\rm (eff)}} \equiv \frac{\Theta_{12}^{(R)}(i\xi_n)}{\Theta_{11}^{(R)}(i\xi_n)}, \quad 
\bar{\Delta}_{W B_1}^{{\rm (eff)}} \equiv -\frac{\Theta_{21}^{(L)}(i\xi_n)}{\Theta_{11}^{(L)}(i\xi_n)}.
\ee
In the weak anisotropy regime, we find that $\bar{\Delta}_{WR}^{{\rm (eff)}}$ and $\bar{\Delta}_{W B_1}^{{\rm (eff)}}$ have the approximate values given by Eqs.~(\ref{eq:victoria}) of App.~\ref{app:multilayers}. 
In the limit of large $N_L$ and $N_R$, these coefficients further simplify to
\begin{subequations}
\ba
\bar{\Delta}_{WR}^{{\rm (eff)}} 
&\approx& 
-\frac{\gamma_n}{2}e^{-Q b'}\sinh(Q b')\cos^2\psi 
\nonumber\\
&&- \frac{\gamma_n e^{-Q b'} \sinh(Q b')}{8(\cosh(2Q(b+b'))-\cos(2\delta\theta))}
\nonumber\\
&&\quad\times 
\big[ 
\cos(2(\delta\theta-\psi)) 
- e^{-2Q(b+b')}\cos(2\psi)
\big]
\nonumber\\
&&-\frac{\gamma_n e^{-Q(b+2b')} \sinh(Q b')}{8 \sinh(Q(b+b'))} 
\ea
\ba
\bar{\Delta}_{W B_1}^{{\rm (eff)}} 
&\approx& 
-\frac{\gamma_n}{2} e^{-Q a'} \sinh(Q a') \cos^2(\theta_d-\psi)
\nonumber\\
&&-
\frac{\gamma_n e^{-Q a'} \sinh(Q a')}{8(\cosh(2Q(a+a'))-\cos(2\delta\theta))}
\nonumber\\
&&\quad\times 
\big[ 
\cos(2(\delta\theta-\theta_d+\psi)) 
\nonumber\\
&&\qquad- 
e^{-2Q(a+a')} \cos(2(\theta_d-\psi))
\big]
\nonumber\\
&&- 
\frac{\gamma_n e^{-Q(a+2a')} \sinh(Q a')}{8 \sinh(Q(a+a'))}.
\ea
\end{subequations}

\subsection{van der Waals interaction free energy}

The interaction free energy per unit area of the interacting multilayered slabs is given by
\ba
\label{eq:Gmulti}
G_{rr} &=& \frac{k_{\rm{B}}T}{4\pi^2} \! \sum_n^\prime \! \int \! dQ\,Q \! \int \! d\psi \,
\ln (1 - \bar{\Delta}_{W B_1}^{(\rm{eff})} \bar{\Delta}_{WR}^{(\rm{eff})} e^{-2Qd})
\nonumber\\
&\approx&
- \frac{k_{\rm{B}}T}{4\pi^2} \! \sum_n^\prime \! \int \! dQ\,Q \int \! d\psi \, 
\bar{\Delta}_{W B_1}^{(\rm{eff})} \bar{\Delta}_{WR}^{(\rm{eff})} e^{-2Qd}
\nonumber\\
&=&
- \frac{\gamma^2 k_{\rm{B}}T}{8\pi^2} \! \int \! dQ\,Q\,e^{-2Q d}
\big[
h_0 + h_1 \cos(2\theta_d) 
\nonumber\\
&&\quad+ h_2 \cos(2(\delta\theta-\theta_d)) 
+ 
h_3 \cos(2(2\delta\theta-\theta_d)) 
\big],
\nonumber\\
\ea
where as before $\gamma^2 \equiv 2\sum_n^\prime \gamma_n^2$, and $h_0$, $h_1$, $h_2$ and $h_3$ are given by Eqs.~(\ref{eq:endrina}) of App.~\ref{app:multilayers}. 

For the case where the optic axes within each multilayer are all aligned (i.e., $\delta\theta=0$), the free energy per unit area simplifies to
\be
G_{mm} \equiv G_{rr}(\delta\theta=0) =
 -\frac{\gamma^2 k_{{\rm B}}T}{512\pi} (1+2\cos^2\theta_d) \, J,
\label{eq:Grr0}
\ee
with
\begin{subequations} 
\ba
\label{eq:J}
J &\equiv& \int_0^\infty\!\!\! dQ \, Q \, e^{-2Q d} f(Q), 
\\
f(Q) &\equiv& \frac{e^{Q(a+b)} \sinh(Q a') \sinh(Q b')}{\sinh(Q(a+a'))\sinh(Q(b+b'))}. 
\label{eq:fQ}
\ea
\end{subequations}
To find the asymptotic behavior of $G_{mm}$, we note that as $Q \rightarrow \infty$, $f(Q) \rightarrow 1$, and as $Q \rightarrow 0$, $f(Q) \rightarrow a' b' /((a+a')(b+b'))$, i.e., $f(Q)$ is always of the order of unity (assuming that the thicknesses $a$, $a'$, $b$ and $b'$ are of comparable order). The variation of the integrand of $J$ is thus determined by the behavior of $Q \, \exp(-2Q d)$, which is peaked at $Q = 1/2d$. The dominant contribution to the integral $J$ thus comes from the modes with $Q \sim 1/2d$. 

Hence for large $d$, the dominant mode contribution comes from $Q \sim 0$, where $J$ can be approximated by 
\be
J \approx \int_0^\infty\!\!\! dQ \, Q \frac{e^{-2Q d} a' b'}{(a+a')(b+b')} = \frac{a' b'}{4(a+a')(b+b')d^2}, 
\ee 
and thus 
\be
G_{mm} \approx  -\frac{\gamma^2 k_{{\rm B}}T (1+2\cos^2\theta_d) a' b'}{2048\pi (a+a')(b+b') d^2}, 
\ee
which corresponds to a force per unit area that decays with $d^{-3}$, viz., 
\be
\mathcal{F}_{mm} \approx  -\frac{\gamma^2 k_{{\rm B}}T (1+2\cos^2\theta_d) a' b'}{1024\pi (a+a')(b+b') d^3}. 
\ee
In the other limit where $d$ is much smaller than the layer thicknesses, $f(Q)$ is dominated by $Q \sim 1/2d$, and thus we can approximate
\be
f(Q) \approx  \frac{e^{\frac{a+b}{2d}} \sinh(\frac{a'}{2d}) \sinh(\frac{b'}{2d})}{\sinh(\frac{a+a'}{2d})\sinh(\frac{b+b'}{2d})} \approx 1, \,\, J \approx \frac{1}{4 d^2},
\ee
and thus 
\be
G_{mm} \approx -\frac{\gamma^2 k_{{\rm B}}T (1+2\cos^2\theta_d)}{2048\pi d^2}. 
\ee
This also corresponds to a force per unit area that decays with $d^{-3}$, which is given by
\be
\mathcal{F}_{mm} \approx -\frac{\gamma^2 k_{{\rm B}}T (1+2\cos^2\theta_d)}{1024\pi d^3}. 
\ee
Thus in both the large and small $d$ limits, $G_{mm}$ decays as $d^{-2}$, and in the small $d$ limit, $G_{mm}$ approximates to the value of $G_{ss}$, which is the free energy per unit area of two thick anisotropic slabs, which is physically reasonable. 

The behavior of the free energy per unit area for two interacting multilayers is shown in Fig.~\ref{fig:rr}. We see that the attraction is strongest when $\delta\theta=0$, and weakest when $\delta\theta=\pi/2$. Comparing with the analogous curves in Fig.~\ref{fig:manybody}, we also note that the vdW attraction for two interacting multilayers is stronger than that for a single anisotropic layer interacting with a multilayer. This behavior is consistent with the idea that additional anisotropic layers (in the presence of a multilayer) increase the extent over which dielectric contrast occurs, thus contributing to an increase in the vdW attraction. 

\subsection{van der Waals torque}

From Eq.~(\ref{eq:Gmulti}) we obtain the vdW torque per unit area for two interacting multi-layered slabs, $\tau_{mm}$:
\ba
\label{eq:trr}
\tau_{rr} &=& 
-\frac{k_{{\rm B}} T}{4\pi^2} \!    
\int\! dQ\, Q\, e^{-2Q d} \big[
h_1 \sin(2\theta_d) 
\\
&&\quad+ 
h_2 \sin(2(\theta_d-\delta\theta)) + h_3 \sin(2(\theta_d-2\delta\theta))
\big].
\nonumber
\ea
For the case where the optic axes within each multilayer are all aligned (i.e., $\delta\theta=0$), the vdW torque per unit area simplifies to
\be
\tau_{mm} \equiv \tau_{rr}(\delta\theta=0) 
= -\frac{\partial G_{mm}}{\partial \theta_d} 
\approx - \frac{\gamma^2 k_{{\rm B}}T \sin(2\theta_d) J}{256 \pi},
\ee
where $J$ is defined by Eq.~(\ref{eq:J}). Similar to the case of $G_{mm}$, we can again determine the asymptotic behavior of $\tau_{mm}$. For large $d$, we find 
\be
\tau_{mm}   
\approx 
- \frac{\gamma^2 k_{{\rm B}}T \sin(2\theta_d) a' b'}{1024 \pi (a+a')(b+b') d^2},
\ee
whereas for small $d$, we find 
\be
\tau_{mm}   
\approx 
- \frac{\gamma^2 k_{{\rm B}}T \sin(2\theta_d)}{1024 \pi d^2}.
\ee
The behavior of the vdW torque is shown in Fig.~\ref{fig:rr}. Analogous to what we have observed for the vdW torque between a single anisotropic layer and a multilayer in Sec.~\ref{sec:vdwtorquesr}, the vdW torque between two multilayers is also strongest for $\delta\theta=0$ and weakest for $\delta\theta=\pi/2$, and such behavior can be similarly understood using the qualitative explanations given in that section. If we compare with the curves for the vdW torque in Fig.~\ref{fig:manybody}, we see that the vdW torque for two multilayers is enhanced relative to that for a single layer interacting with a multilayer if $\delta\theta=0$, and relatively reduced if $\delta\theta=\pi/2$. Finally, we note that the ``phase shift" is more pronounced for the case of two multilayers.

\section{Optic axis perpendicular to plane of anisotropic layers}

We now turn our attention to the case where the optic axis of each dielectrically anisotropic, uniaxial layer is \emph{perpendicular} (rather than parallel) to the plane of the layer, and the layers are ``stacked" co-axially as before. In this case the optic axes of the anisotropic layers are all parallel, and there is no vdW torque. The dielectric tensor in diagonal form is given by 
\be
\bm{\varepsilon}_{B'}^{(\rm{prin})} = 
\begin{pmatrix}
    \varepsilon_{\bing{\perp}}      & 0 & 0 \\
    0 & \varepsilon_{\bing{\perp}} & 0 \\
    0 & 0 & \varepsilon_{\bing{||}}
\end{pmatrix},
\label{eq:anisot}
\ee
where $\varepsilon_{\bing{\perp}}$ ($\varepsilon_{\bing{||}}$) is the dielectric permittivity in a direction perpendicular to (parallel with) the optic axis, and $n$ labels the Matsubara frequencies, $\xi_n = 2\pi k_{{\rm B}}T n/\hbar$. 

As in Sec~\ref{sec:free_energy} we start from the Laplace equation Eq.~(\ref{eq:maxwell1}) with Eq.~(\ref{eq:anisot}), obtaining Eq.~(\ref{eq:f}) with $\rho_{i} = \sqrt{u^2+v^2} \equiv Q$ if layer $i$ is \bing{the solvent}, and $\rho_{i} = \rho_{B'}$ if the layer is $B'$-type, where $\rho_{B'}$ is now given by
\be
\rho_{B'} \equiv \sqrt{\frac{\varepsilon_{\bing{\perp}}}{\varepsilon_{\bing{||}}}} Q 
\ee 
The reflection coefficient for the dielectric discontinuity at the \bing{solvent}-$B'$-type interface can be found from Eq.~(\ref{eq:Delta}), where now 
\ba
\bar{\Delta}_{WB'} &=& 
\frac{\varepsilon_{W} Q - \varepsilon_{\bing{||}}\rho_{B'}}{\varepsilon_{W} Q + \varepsilon_{\bing{||}}\rho_{B'}} 
\nonumber\\
&=& 
\frac{1-g_n(\varepsilon_{\bing{\perp}}/\varepsilon_{W})}{1+g_n(\varepsilon_{\bing{\perp}}/\varepsilon_{W})}
\nonumber\\
&\equiv& \bar{\Delta}_n, 
\ea
where $g_n \equiv (\varepsilon_{\bing{||}}/\varepsilon_{\bing{\perp}})^{1/2} = (\varepsilon_{B'zz}/\varepsilon_{B'xx})^{1/2}$. Similar to Ref.~\cite{parsegian-weiss}, we make the simplifying assumption that $\varepsilon_{\bing{\perp}, n} = \varepsilon_{W}$, from which we obtain 
\be
\bar{\Delta}_n = \frac{1-g_n}{1+g_n}.
\ee
We can write $g_n = 1 +\delta g_n$, where $\delta g_n \equiv (\varepsilon_{\bing{||}}/\varepsilon_{\bing{\perp}})^{1/2}-1$ is a measure of dielectric anisotropy. For weak anisotropy ($\delta g_n \ll 1$), $\bar{\Delta}_n$ can be approximated by 
\be
\bar{\Delta}_n \approx -\frac{1}{2}\delta g_n.
\ee

\subsection{Two interacting single layers}
We first consider the case of two single uniaxial layers of the same thickness $b'$, interacting across a \bing{solvent} layer of thickness $d$. In this case the analogue of Eq.~(\ref{eq:Theta_ss}) is given by
\be
\label{eq:Theta_ssp}
\mathbf{\Theta}^{(\rm{ss})} \equiv \mathbf{D}_{WB'}  \mathbf{T}_{B'} 
\mathbf{D}_{B'W}  \mathbf{T}_{m} \mathbf{D}_{WB'} \mathbf{T}_{B'} \mathbf{D}_{B'W} 
\ee
In the above the matrices are given by 
\be
\mathbf{D}_{WB'} 
\equiv
\begin{pmatrix}
1 & -\bar{\Delta}_{n} \\
-\bar{\Delta}_{n} & 1
\end{pmatrix},
\ee
where 
\be
\mathbf{T}_{B'} 
\equiv 
\begin{pmatrix}
1 & 0 \\
0 & e^{-2 Q b' g_n^{-1}}
\end{pmatrix}, 
\,
\mathbf{T}_{m} 
\equiv 
\begin{pmatrix}
1 & 0 \\
0 & e^{-2 Q d}
\end{pmatrix}. 
\ee   
We find
\be
\Theta_{11}^{(\rm{ss})}(i\xi_n) = (1-\bar{\Delta}_n^2 e^{-2Q b'/g_n})^2 - \bar{\Delta}_n^2 (1-e^{-2Q b'/g_n})^2 e^{-2Q d}
\ee
The free energy per unit area is given by 
\be
G_\perp = \frac{k_{\rm{B}}T}{2\pi} \! \sum_{n}^\prime \! \int_0^\infty \!\!\! dQ \, Q
\ln \frac{\Theta_{11}^{({\rm ss})} (d, i\xi_n)}{\Theta_{11}^{({\rm ss})} (d\rightarrow\infty, i\xi_n)}. 
\ee
Specializing to the weak anisotropy regime, i.e., $\delta g_n \ll 1$, we can approximate the free energy per unit area by
\be
G_\perp \approx
-\frac{\delta g^2 k_{{\rm B}}T}{64\pi} \left[ \frac{1}{d^2} - \frac{2}{(d+b')^2} + \frac{1}{(d+2b')^2} \right], 
\label{eq:Gssperp}
\ee
where $\delta g^2 \equiv 2\sum_n^\prime \delta g_n^2$, and $\sum_{n}^{\prime}$ is a sum running from $n=0$ to $n=\infty$, but with the $n=0$ multiplied by an additional factor of $1/2$. The corresponding force per unit area is given by 
\be
\mathcal{F}_\perp \approx
-\frac{\delta g^2 k_{{\rm B}}T}{32\pi} \left[ \frac{1}{d^3} - \frac{2}{(d+b')^3} + \frac{1}{(d+2b')^3} \right]. 
\ee
The decay behavior of $G_{\perp}$ is essentially the same as that of two single layers with optic axes parallel to the plane of the layers (cf. Eq.~(\ref{eq:Gss})). 

If we consider the high temperature limit, such that $i\xi_n \rightarrow 0$ (and $\varepsilon_{||}, \varepsilon_{\perp} \rightarrow 1$) for $n \neq 0$, then we can make a convenient comparison with the case of two interacting single anisotropic layers whose optic axes lie in the plane of the layers, viz., Eq.~(\ref{eq:Gss_app}). Let us write $\alpha \equiv (\varepsilon_{{\rm opt}}/\varepsilon_{{\rm nopt}})^{1/2}$, where $\varepsilon_{{\rm opt}}$ ($\varepsilon_{{\rm nopt}}$) denotes the principal dielectric permittivity along (perpendicular to) the optic axis, so that in the limit of weak anisotropy, $\alpha \approx 1$. For a system where the optic axis is parallel to the $x$-axis of (and parallel to the plane of) the reference $B'$-type layer, $\alpha = (\varepsilon_{B' xx,0}/\varepsilon_{B' zz,0})^{1/2}$, and $\gamma^2 \rightarrow \gamma_0^2 = (\alpha^2-1)^2 \approx 4(\alpha-1)^2$. On the other hand, for a system where the optic axis is perpendicular to the plane of the layer, which we take to be parallel to the $z$-axis, $\alpha = (\varepsilon_{B' zz,0}/\varepsilon_{B' xx,0})^{1/2}$, and $\delta g^2 \rightarrow \delta g_0^2 = (\alpha-1)^2$. 
Taking $G_{ss}$ from Eq.~(\ref{eq:Gss_app}) for the former system and $G_{\perp}$ from Eq.~(\ref{eq:Gssperp}) for the latter system, we find 
\be
\frac{G_{\perp}}{G_{ss}} \rightarrow \frac{32\delta g_0^2}{\gamma_0^2 (1+2\cos^2\theta_d)} = \frac{8}{1+2\cos^2\theta_d}
\label{eq:compareperpss}
\ee
For weak anisotropy and at high temperature, two single anisotropic layers thus attract each other more strongly when their optic axes are oriented perpendicular to the plane of the layers than when the optic axes are parallel to the plane. 

\subsection{Single layer interacting with multilayer}
Next, we consider a single anisotropic layer interacting with a stack of $N+1$ anisotropic layers across an intervening isotropic medium. The matrices $\mathbf{\Theta}^{({\rm sm})}$ and $\mathbf{A}$ are still expressed by the formulas Eqs.~(\ref{eq:Theta_sm}) and (\ref{eq:A}), where now the matrices are given by
\begin{subequations}
\label{eq:matrixes}
\ba
\mathbf{D}_{WB'} &=& \mathbf{D}_{W B_1} 
\equiv
\begin{pmatrix}
1 & -\bar{\Delta}_n \\
-\bar{\Delta}_n & 1
\end{pmatrix}, 
\\
\mathbf{D}_{B'W} &=& \mathbf{D}_{B_1 W} 
\equiv
\begin{pmatrix}
1 & \bar{\Delta}_n \\
\bar{\Delta}_n & 1
\end{pmatrix},
\\
\mathbf{T}_{B} 
&\equiv& 
\begin{pmatrix}
1 & 0 \\
0 & e^{-2 Q b}
\end{pmatrix}, 
\\
\mathbf{T}_{B'} &=& \mathbf{T}_{B_1} 
\equiv
\begin{pmatrix}
1 & 0 \\
0 & e^{-2 Q b' g_n^{-1}}
\end{pmatrix},
\\
\mathbf{T}_{m} 
&\equiv& 
\begin{pmatrix}
1 & 0 \\
0 & e^{-2 Q d}
\end{pmatrix}.
\ea
\end{subequations}
The elements of the matrix $\mathbf{A}_n$ are given by
\begin{subequations}
\label{eq:elementsAperp}
\ba
A_{11} &=& 1 - e^{-2Q b' g_n^{-1}} \bar{\Delta}_{n}^2,
\\
A_{12} &=& e^{-2Q b} (1 - e^{-2Q b' g_n^{-1}}) \bar{\Delta}_{n}, 
\\
A_{21} &=& - (1 - e^{-2Q b' g_n^{-1}}) \bar{\Delta}_{n},
\\
A_{22} &=& e^{-2Q b}(e^{-2Q b' g_n^{-1}} - \bar{\Delta}_{n}^2)
\ea 
\end{subequations}
and the determinant is 
\be
|\mathbf{A}| = (1-\bar{\Delta}_{n}^2)^2 e^{-2Q(b+b' g_n^{-1})}
\ee
The matrix product $\mathbf{A}^N$ can be found using Abel\`{e}s' formula~(see, e.g., Ref.~\cite{parsegian-vdw}), which gives
\be
\label{eq:AN_matrix}
\mathbf{A}^{N} = 
\begin{pmatrix}
A_{11}^{(N)} & A_{12}^{(N)} \\
A_{21}^{(N)} & A_{22}^{(N)}
\end{pmatrix}
\ee
where 
\begin{subequations}
\label{eq:AN_matrix_coeffs}
\ba
A_{11}^{(N)} &\equiv& \bigg( \frac{A_{11}}{\sqrt{|\mathbf{A}|}} U_{N-1} - U_{N-2} \bigg) |\mathbf{A}|^{N/2},
\\
A_{12}^{(N)} &\equiv& A_{12} U_{N-1} |\mathbf{A}|^{(N-1)/2},
\\
A_{21}^{(N)} &\equiv& A_{21} U_{N-1} |\mathbf{A}|^{(N-1)/2},
\\
A_{22}^{(N)} &\equiv& \bigg( \frac{A_{22}}{\sqrt{|\mathbf{A}|}} U_{N-1} - U_{N-2} \bigg) |\mathbf{A}|^{N/2}
\ea
\end{subequations}
Here $U_N$ are the Chebyshev polynomials that we already encountered in Eq.~(\ref{eq:chebyshev}). 
For weak anisotropy, we can approximate $\xi$ (defined in Eq.~(\ref{eq:xi_def})) by 
\be
\xi \approx Q(b+b') - \delta g_n \, Q b'.
\ee
We can compute $\Theta_{11}^{{\rm (sm)}}$ from Eq.~(\ref{eq:Theta_sm}). For weak anisotropy ($\bar{\Delta}_n \approx -\delta g_n/2 \ll 1$) we find
\ba
&&\frac{\Theta_{11}^{{\rm (sm)}}(d, i\xi_n)}{\Theta_{11}^{{\rm (sm)}}(d \rightarrow \infty, i\xi_n)} 
\\
&\!\!\approx\!\!& 
1 - \frac{\delta g_n^2 \sinh^2(Q b') \, e^{-2Q b'-2Q d}}{1-\frac{\sinh((N-1)Q(b+b'))}{\sinh(NQ(b+b'))}e^{-Q(b+b')}} 
\nonumber\\
&&\times \! 
\left[  
1 + e^{-2Q(b+b')} - \frac{e^{-Q(b+b')} \sinh((N-1)Q(b+b'))}{\sinh(NQ(b+b'))} 
\right]
\nonumber
\ea
In the limit of large $N$, the above simplifies to
\be
\frac{\Theta_{11}^{{\rm (sm)}}(d, i\xi_n)}{\Theta_{11}^{{\rm (sm)}}(d \rightarrow \infty, i\xi_n)} 
\approx 
1 - \frac{\delta g_n^2 \sinh^2(Q b') \, e^{-2Q b' - 2Q d}}{1-e^{-2Q(b+b')}}.
\ee
In the regime of weak anisotropy and for large $N$, the interaction free energy per unit area is thus given by
\ba
G_\perp &=& 
\frac{k_{\rm{B}}T}{2\pi} \! \sum_{n}^{\prime} 
\int_0^\infty \!\!\!\!dQ \, Q \,
\ln \frac{\Theta_{11}^{{\rm (sm)}}(d, i\xi_n)}{\Theta_{11}^{{\rm (sm)}}(d \rightarrow \infty, i\xi_n)} 
\nonumber\\
&\approx&
-\frac{\delta g^2 \, k_{{\rm B}}T}{64 \pi (b+b')^2} 
\Big[ \psi^{(1)}\left( \frac{d}{b+b'} \right) 
\nonumber\\
&&-
2 \psi^{(1)}\left( \frac{d+b'}{b+b'} \right) + \psi^{(1)}\left( \frac{d+2b'}{b+b'} \right) \Big],
\label{eq:Gsm_z}
\ea
where $\sum_{n}^{\prime}$ is a sum running from $n=0$ to $n=\infty$, but with the $n=0$ multiplied by an additional factor of $1/2$, and $\delta g^2 \equiv 2\sum_{n}^{\prime} \delta g_n^2$. The corresponding force per unit area is given by 
\ba
\mathcal{F}_\perp 
&\approx&
\frac{\delta g^2 \, k_{{\rm B}}T}{64 \pi (b+b')^3} 
\Big[ \psi^{(2)}\left( \frac{d}{b+b'} \right) 
\nonumber\\
&&-
2 \psi^{(2)}\left( \frac{d+b'}{b+b'} \right) + \psi^{(2)}\left( \frac{d+2b'}{b+b'} \right) \Big],
\ea
where $\psi^{(2)}(z)$ is the second derivative of the digamma function $\psi(z)$. The decay behavior of $G_{\perp}$ is qualitatively the same as that for the corresponding system with optic axes all parallel to the plane of the layers, in which the optic axes in the multilayer are all aligned (cf. Eq.~(\ref{eq:Gsm2})).    

In the high temperature limit we can compare the strengths of the van der Waals attraction for the cases of optic axes aligned parallel to the plane of the anisotropic layers, viz., $G_{sm}$ from Eq.~(\ref{eq:Gsm2}), and optic axes aligned perpendicular to the plane of the layers, viz., $G_{\perp}$ from Eq.~(\ref{eq:Gsm_z}). In this case, defining again $\alpha \equiv (\varepsilon_{{\rm opt}}/\varepsilon_{{\rm nopt}})^{1/2}$, we have $\gamma^2 \rightarrow \gamma_0^2 \approx 4(\alpha-1)^2$ and $\delta g^2 \rightarrow \delta g_0^2 = (\alpha-1)^2$. We find 
\be
\label{eq:compareperpsr}
\frac{G_\perp}{G_{sm}} \rightarrow \frac{32\delta g_0^2}{\gamma_0^2 (1+2\cos^2\theta_d)} = \frac{8}{1+2\cos^2\theta_d},
\ee
i.e., the same value that we found in Eq.~(\ref{eq:compareperpss}) for the case of two interacting single anisotropic layers. 

\subsection{Two interacting multilayers}
For the case of two interacting multi-layered slabs separated by a \bing{solvent} layer of thickness $d$, the corresponding transfer matrix is given by 
\be
\mathbf{\Theta}^{({\rm mm})} = \mathbf{\Theta}^{(R)}\cdot\mathbf{T}_m \cdot \mathbf{\Theta}^{(L)},
\ee
where 
\begin{subequations}
\ba
\mathbf{\Theta}^{(R)} &\equiv& \mathbf{A}^{N_R} \mathbf{D}_{WB'} \mathbf{T}_{B'} \mathbf{D}_{B'W},
\\
\mathbf{\Theta}^{(L)} &\equiv& \mathbf{D}_{WB'} \widetilde{\mathbf{T}}_{B'} \mathbf{D}_{B'W} \widetilde{\mathbf{A}}^{N_L},
\\
\mathbf{A} &\equiv& \mathbf{D}_{WB'}\mathbf{T}_{B'}\mathbf{D}_{B'W}\mathbf{T}_{B},
\\
\widetilde{\mathbf{A}} &\equiv& \widetilde{\mathbf{T}}_B \mathbf{D}_{WB'} \widetilde{\mathbf{T}}_{B'} \mathbf{D}_{B'W}
\ea
\end{subequations}
The matrices $\mathbf{D}_{WB'}$, $\mathbf{D}_{B'W}$, $\mathbf{T}_{B}$, $\mathbf{T}_{B'}$ and $\mathbf{T}_{m}$ are introduced in Eqs.~(\ref{eq:matrixes}), and the matrices $\widetilde{\mathbf{T}}_{B}$ and $\widetilde{\mathbf{T}}_{B'}$ are defined by 
\begin{subequations}
\ba
\widetilde{\mathbf{T}}_{B} 
&\equiv& 
\begin{pmatrix}
1 & 0 \\
0 & e^{-2 Q a}
\end{pmatrix}, 
\\
\widetilde{\mathbf{T}}_{B'} 
&\equiv& 
\begin{pmatrix}
1 & 0 \\
0 & e^{-2 Q a' g_n^{-1}}
\end{pmatrix} 
\ea 
\end{subequations}
The elements of matrix $\mathbf{A}$ are given in Eqs.~(\ref{eq:elementsAperp}), whilst those of matrix $\widetilde{\mathbf{A}}$ are given by 
\begin{subequations} 
\ba
\widetilde{A}_{11} &=& 1 - \bar{\Delta}_n^2 e^{-2Q a' g_n^{-1}};
\\
\widetilde{A}_{12} &=& \bar{\Delta}_n (1 - e^{-2Q a' g_n^{-1}});
\\
\widetilde{A}_{21} &=& -\bar{\Delta}_n (1 - e^{-2Q a' g_n^{-1}}) e^{-2Q a};
\\
\widetilde{A}_{22} &=& (e^{-2Q a' g_n^{-1}} - \bar{\Delta}_n^2) e^{-2Q a}
\ea
\end{subequations}
The corresponding determinant is 
\be
|\widetilde{\mathbf{A}}| = (1 - \bar{\Delta}_n^2)^2 e^{-2Q (a+a' g_n^{-1})}.
\ee
As in Sec.~\ref{sec:rr}, the matrix element $\Theta_{11}$ is given by
\be
\Theta_{11}^{({\rm mm})} = \Theta_{11}^{(L)} \Theta_{11}^{(R)} + \Theta_{21}^{(L)} \Theta_{12}^{(R)} e^{-2Q d},
\ee
from which we deduce
\be
\frac{\Theta_{11}^{({{\rm mm}})}(d,i\xi_n)}{\Theta_{11}^{({{\rm mm}})}(d\rightarrow\infty,i\xi_n)} 
= 1 - \bar{\Delta}_{WR}^{{\rm (eff)}}(i\xi_n) \bar{\Delta}_{W B_1}^{{\rm (eff)}}(i\xi_n) e^{-2Q d},
\ee
where $\bar{\Delta}_{W B_1}^{\rm{(eff)}}$ and $\bar{\Delta}_{WR}^{\rm{(eff)}}$ are the effective reflection coefficients for the left and right slabs, defined by 
\be
\bar{\Delta}_{WR}^{{\rm (eff)}} \equiv \frac{\Theta_{12}^{(R)}}{\Theta_{11}^{(R)}}, \quad 
\bar{\Delta}_{W B_1}^{{\rm (eff)}} \equiv -\frac{\Theta_{21}^{(L)}}{\Theta_{11}^{(L)}}.
\ee
For weak anisotropy, these coefficients can be approximated by
\begin{subequations}
\ba
&&\bar{\Delta}_{WR}^{{\rm (eff)}} 
\approx
\nonumber\\
&&-\frac{\delta g_n e^{-Q b'}\sinh(Q b')}{1-\frac{e^{-Q(b+b')} \sinh((N_R-1)Q(b+b'))}{\sinh(N_R Q(b+b'))}}
\bigg[ 
1 + e^{-2Q(b+b')} 
\nonumber\\
&&
\quad
- \frac{e^{-Q(b+b')} \sinh((N_R-1)Q(b+b'))}{\sinh(N_R Q(b+b'))}
\bigg],
\\
&&\bar{\Delta}_{W B_1}^{{\rm (eff)}} 
\approx
\nonumber\\
&&-\frac{\delta g_n e^{-Q a'}\sinh(Q a')}{1-\frac{e^{-Q(a+a')} \sinh((N_L-1)Q(a+a'))}{\sinh(N_L Q(a+a'))}}
\bigg[ 
1 + e^{-2Q(a+a')} 
\nonumber\\
&&
\quad
- \frac{e^{-Q(a+a')} \sinh((N_L-1)Q(a+a'))}{\sinh(N_L Q(a+a'))}
\bigg].
\ea
\end{subequations}
For large $N_L$ and $N_R$, we can approximate
\begin{subequations}
\ba
&&\bar{\Delta}_{WR}^{{\rm (eff)}} 
\approx -\frac{\delta g_n e^{Q b} \sinh(Q b')}{2\sinh(Q(b+b'))}, 
\\
&&\bar{\Delta}_{W B_1}^{{\rm (eff)}} 
\approx -\frac{\delta g_n e^{Q a} \sinh(Q a')}{2\sinh(Q(a+a'))}.
\ea
\end{subequations}
The interaction free energy per unit area is given by
\ba
\label{eq:Gmm_z}
&&G_\perp 
\\
&\!\!=\!\!& 
\frac{k_{\rm{B}}T}{2\pi} \! \sum_{n}^{\prime} 
\int_0^\infty \!\!\!\!dQ \, Q \,
\ln(1 - \bar{\Delta}_{W B_1}^{\rm{(eff)}} \bar{\Delta}_{WR}^{\rm{(eff)}} e^{-2Qd})
\nonumber\\
&\!\!\approx\!\!& 
-\frac{\delta g^2 k_{\rm{B}}T}{16\pi} \!
\int_0^\infty \!\!\!\!dQ \, Q \,
\frac{e^{Q(a+b)-2Qd} \sinh(Qa')\sinh(Qb')}{\sinh(Q(a+a'))\sinh(Q(b+b'))}.
\nonumber
\ea
Up to a prefactor this is the same free energy expression as that for two interacting multilayers with optic axes parallel to the plane of the layers, in which the optic axes in a given multilayer are all aligned (cf. Eq.~(\ref{eq:Grr0})). Thus using similar arguments $G_\perp$ decays with $d^{-2}$, and the corresponding force $\mathcal{F}_\perp$ decays with $d^{-3}$. 

For the case of high temperature, we can again compare the van der Waals interaction strengths for the cases of optic axes aligned perpendicular to the plane of the anisotropic layers, i.e., $G_\perp$ from Eq.~(\ref{eq:Gmm_z}), and optic axes aligned parallel to the plane of the layers, viz., $G_{mm}$ from Eq.~(\ref{eq:Grr0}). Using $\alpha \equiv (\varepsilon_{{\rm opt}}/\varepsilon_{{\rm nopt}})^{1/2}$, $\gamma^2 \rightarrow \gamma_0^2 \approx 4(\alpha-1)^2$ and $\delta g^2 \rightarrow \delta g_0^2 = (\alpha-1)^2$, we find 
\be
\frac{G_\perp}{G_{mm}} \rightarrow \frac{32\delta g_0^2}{\gamma_0^2 (1+2\cos^2\theta_d)} = \frac{8}{1+2\cos^2\theta_d},
\ee
which is the same result that we found for the case of two interacting single anisotropic layers, Eq.~(\ref{eq:compareperpss}), and the case of a single anisotropic layer interacting with an anisotropic multilayer whose optic axes are all aligned, Eq.~(\ref{eq:compareperpsr}). 

\section{Summary and conclusion}

In this Paper we have studied the behavior of the van der Waals (vdW) torque and interaction free energy of dielectrically anisotropic layered media, in the regime of \bing{weak dielectric anisotropy, no retardation, and where the dielectric coefficients of the ordinary axes of the uniaxial crystal layers match the dielectric permittivity of the solvent medium.} In particular we have examined the behavior of the following three systems: (i)~two interacting single anisotropic layers, (ii)~a single anisotropic layer interacting with an anisotropic multilayered slab, and (iii)~two interacting anisotropic multilayered slabs. We have considered these systems in the following two cases: (a)~the optic axes lie in the plane of the layers, and (b)~the optic axes are perpendicular to the plane of the layers. For case (a), we have considered two further scenarios, one where all the optic axes of the anisotropic layers in a given multilayer are aligned, and the other where the optic axes undergo constant angular increments $\delta\theta$ across the multilayer. 

We summarize our results for case~(a) as follows. We found that increasing the thicknesses of the anisotropic layers has the effect of increasing the magnitude of the vdW interaction free energy and torque. 
Moreover, we found that the vdW attraction is strongest for two interacting multilayers and weakest for two interacting single anisotropic layers, for all values of $\delta\theta$. On the other hand, the amplitude of the vdW torque is largest for two multilayers and smallest for two single layers when $\delta\theta=0$, but the torque amplitude is smallest for two multilayers and largest for two single layers when $\delta\theta=\pi/2$. In addition, the angle $\theta_d$ (the relative orientation between the optic axes of the oppositely facing anisotropic layers of the two interacting layered media) at which the layered media are in a stable (unstable) configuration of zero overall torque is an integer (half-integer) factor of $\pi$ for $\delta\theta=0$ and $\delta\theta=\pi/2$, but moves away from these values as we tune $\delta\theta$ from $0$ to $\pi/2$.  We have also determined the asymptotic behaviors of the vdW free energy and torque for the three systems in the case where $\delta\theta=0$. For separations that are much larger than the thickness of each anisotropic layer, we found that the free energy and torque decay as  $d^{-4}$ in the case of system (i), $d^{-3}$ in the case of (ii), and $d^{-2}$ in the case of (iii). On the other hand, if the separation is much smaller than the layer thicknesses, the free energy and torque of all three systems approach those corresponding to two very thick anisotropic layers, decaying with $d^{-2}$. 

For case (b) (optic axes directed perpendicular to the plane of the layers), we have found that the free energies have the same decay behaviors as those in case (a). In the high temperature limit, we have found that the free energies for (b) are larger than those for (a) by the same factor, viz., $8/(1+2\cos^2\theta_d)$. 

Although the vdW torque has been analysed and calculated in different setups, a direct experiment - though in principle feasible - is still sorely lacking \cite{Capasso2,Capasso3}.  We believe that multilayered systems, of the type analysed here, are probably the most straightforward option for an experimental confirmation of this less commonly appreciated feature of vdW interactions. In particular, liquid crystalline arrays of the smectic C* type should prove potentially relevant for this endeavour as they can self-assemble from the solution and their properties can be controlled by macroscopic fields. In smectic C* arrays the director makes a tilt angle with respect to the smectic layer that furthermore rotates from layer to layer forming a helix, implying furthermore also rotating optic axes that could be controlled by temperature or other external fields and fine tuned for the different experimental setups. This multilayer configuration would be in addition directly describable by the formalism derived and developed above.

The approach and analysis described in this work opens up further possible avenues of investigation. An obvious extension would be to include the effects of retardation, though formally this could be quite demanding \cite{Veble2} as even two semi-infinite layers lead to very unwielding formulae \cite{barash1,barash2,Philbin}. Another line of inquiry is to explore the effects of random disorder in the alignments of the optic axes on the vdW torque and interaction in general, along similar lines as for the case of a disordered isotropic dielectric function \cite{Dean}. Yet another area of research can be to study the nanolevitation of plane-parallel multilayers caused by vdW repulsion in real systems that could be controlled by macroscopic external fields that would manipulate the degree of anisotropy.  In all these listed cases the compendium of results described in this work would be of significant value.

\section{Acknowledgments}
BSL thanks J. Munday, J. F. Dobson, B. Guizal, A. Gambassi, and V. Esteso for constructive discussions. The authors acknowledge support from the Slovene Agency of Research and Development (ARRS) through Grant No. P1-0055.

\eject
\newpage

\appendix

\section{The element $\Theta_{11}^{({\rm ss})}$}
\label{app:eli}

Here we show the explicit formula for the element $\Theta_{11}^{({\rm ss})}$:
\ba
\label{eq:Theta11_ss}
&&\Theta_{11}^{({{\rm ss}})} 
\\
&=& 
1 + e^{-2Qb' g_{B_1}} \bar{\Delta}_{B_1 W} \bar{\Delta}_{W B_1} 
\nonumber\\
&&+ 
[e^{-2Q d}  \bar{\Delta}_{B_1 W} + e^{-2Q(d + b' g_{B_1})}  \bar{\Delta}_{W B_1}]   \bar{\Delta}_{W B_2} 
\nonumber\\
&&+ 
\big[
e^{-2Q(d+ b' g_{B_2})} \bar{\Delta}_{B_1 W} 
+ e^{-2Q b' g_{B_2}} \bar{\Delta}_{W B_2}
\nonumber\\
&&\quad
+ e^{-2Q(d+ b' (g_{B_1} + g_{B_2}))} \bar{\Delta}_{W B_1}
\nonumber\\
&&\quad+ 
e^{-2Q(b' (g_{B_1} + g_{B_2}))} 
\nonumber\\
&&\qquad \times 
\bar{\Delta}_{B_1 W} \bar{\Delta}_{W B_2} \bar{\Delta}_{W B_1}
\big]  
\bar{\Delta}_{B_2 W}
\nonumber
\ea

\section{Single anisotropic layer interacting with multilayer having aligned optic axes} 
\label{app:1}

The elements of the matrix $\mathbf{A}$ defined in Eq.~(\ref{eq:A}) are found to be
\begin{subequations}
\label{eq:Asmele}
\ba
A_{11} &=& 1 - e^{-2Q b' g_{B'}} \bar{\Delta}_{WB'}^2,
\\
A_{12} &=& e^{-2Q b} (1 - e^{-2Q b' g_{B'}}) \bar{\Delta}_{WB'},
\\
A_{21} &=& - (1 - e^{-2Q b' g_{B'}}) \bar{\Delta}_{WB'},
\\
A_{22} &=& e^{-2Q b}(e^{-2Q b' g_{B'}} - \bar{\Delta}_{WB'}^2),
\ea 
\end{subequations}
and the determinant is 
\be
|\mathbf{A}| = (1-\bar{\Delta}_{WB'}^2)^2 e^{-2Q(b+b' g_{B'})}
\ee
The coefficients in the numerator of $\bar{\Delta}^{({\rm eff})}$ in Eq.~(\ref{eq:Deff_exact}) are given by
\begin{subequations}
\label{eq:scoeffs}
\ba
s_0 &\!\!\equiv\!\!& - e^{-2 Q b' g_{B'}} A_{12},
\\
s_1 &\!\!\equiv\!\!& 2 ( w \sqrt{|\mathbf{A}|} - A_{11} ) \,
e^{-Q b' g_{B'}} 
\sinh(Q b' g_{B'}),
\\
s_2 &\!\!\equiv\!\!& A_{12} 
\ea
\end{subequations}
and coefficients in the denominator are given by
\begin{subequations}
\label{eq:tcoeffs}
\ba
t_{0} &\equiv& w \sqrt{|\mathbf{A}|} - A_{11}, 
\\
t_{1} &\equiv& 2 A_{12} e^{-Q b' g_{B'}} \sinh(Q b' g_{B'}),
\\
t_{2} &\equiv& ( A_{11} - w \sqrt{|\mathbf{A}|} ) e^{-2Q b' g_{B'}}
\ea
\end{subequations}
In the above $w \equiv U_{N-2}/U_{N-1}$. 

\section{Single anisotropic layer interacting with multilayer having rotating optic axes} 
\label{app:2}
The elements of the matrix $\mathbf{A}^{(j)}$ in Eq.~(\ref{eq:Ai_matrix}) are given below:
\begin{subequations}
\label{eq:Asla}
\ba
A_{11}^{(j)} &=& 1 - e^{-2Q b' g_{B'}^{(j)}} (\bar{\Delta}_{WB'}^{(j)})^2,
\\
A_{12}^{(j)} &=& e^{-2Q b} (1 - e^{-2Q b' g_{B'}^{(j)}}) \bar{\Delta}_{WB'}^{(j)},
\\
A_{21}^{(j)} &=& - (1 - e^{-2Q b' g_{B'}^{(j)}}) \bar{\Delta}_{WB'}^{(j)},
\\
A_{22}^{(j)} &=& e^{-2Q b}(e^{-2Q b' g_{B'}^{(j)}} - (\bar{\Delta}_{WB'}^{(j)})^2)
\ea 
\end{subequations}
and the determinant is 
\be
|\mathbf{A}^{(j)}| = (1-(\bar{\Delta}_{WB'}^{(j)})^2)^2 e^{-2Q(b+b' g_{B'}^{(j)})}
\ee
The matrix elements of $\mathbf{A}_0$, $\delta\mathbf{A}^{(j)}$ and $\mathbf{B}$ introduced Eqs.~(\ref{eq:Ajappr}) and (\ref{eq:B}) are given below: 
\be
A_{0,11} = 1, A_{0,12}=A_{0,21}=0,
A_{0,22} = e^{-2Q(b+b')}
\label{eq:Amatrixelements}
\ee
\begin{subequations}
\label{eq:deltaAele}
\ba
\delta A_{11}^{(j)} &\!=\!& 0,
\\
\delta A_{12}^{(j)} &\!=\!& -\frac{\gamma_n}{2} (\cos(j\delta\theta-\psi))^2 e^{-Q(2b+b')}\sinh(Qb'),
\nonumber\\
\\
\delta A_{21}^{(j)} &\!=\!& \frac{\gamma_n}{2} (\cos(j\delta\theta-\psi))^2 e^{-Qb'}\sinh(Qb'),
\\
\delta A_{22}^{(j)} &\!=\!& -\gamma_n (\cos(j\delta\theta-\psi))^2 Qb' e^{-2Q(b+b')}
\ea
\end{subequations}
\begin{subequations}
\label{eq:Bele}
\ba
B_{11} &=& 0, 
\\
B_{12} &=&
-\frac{\gamma_n}{2} 
e^{Q b' -2NQ(b+b')} \sinh(Qb') 
\nonumber\\ 
&&\quad \times 
V_N(-Q(b+b'), \delta\theta, \psi),
\\
B_{21} &=& 
\frac{\gamma_n}{2} 
e^{-Q b'} \sinh(Q b') \,  
V_N(Q(b+b'), \delta\theta, \psi),
\nonumber\\
\\
B_{22} &=& 
-\frac{\gamma_n}{2} 
Qb' \, e^{-2NQ(b+b')} (N+P_N(\delta\theta))
\ea
\end{subequations}
In the above, the functions $V_N(t,\delta\theta,\psi)$ and $P_N(\delta\theta)$ are defined by
\ba
V_N(t,\delta\theta,\psi) &\equiv& \sum_{j=1}^{N} (\cos(j\delta\theta-\psi))^2 e^{-2(N-j)t}
\nonumber\\
&=& 
\frac{1}{4(\cosh 2t - \cos 2\delta\theta)} \big[ e^{2t} \cos(2(N\delta\theta-\psi))
\nonumber\\
&&-\cos(2((N+1)\delta\theta-\psi)) 
\nonumber\\
&&-e^{-2(N-1)t} \cos2\psi 
\nonumber\\
&&+2e^{-(N-1)t} (\cosh 2t - \cos 2\delta\theta) \frac{\sinh Nt}{\sinh t}
\nonumber\\
&&+e^{-2Nt} \cos(2(\delta\theta-\psi))
\big],
\label{eq:VN}
\\
P_N(\delta\theta) &\equiv&
2\sum_{j=1}^{N} (\cos(j\delta\theta-\psi))^2 - N
\nonumber\\
&=& 
\cos((N+1)\delta\theta - 2\psi) \frac{\sin N\delta\theta}{\sin \delta\theta}
\ea
$V_N$ can be verified e.g. by setting $\delta\theta=0$ and $\psi=0$ and evaluating the sum on the LHS and the formula on the RHS, and seeing that they agree.

The transfer matrix elements in Eqs.~(\ref{eq:Thetasinglemany}) are given by
\begin{subequations}
\label{eq:A1}
\ba
\Theta_{11}^{(R)} &\!\!=\!\!& 1 - B_{12} \bar{\Delta}_{WB'}^{(0)} + (B_{12} - \bar{\Delta}_{WB'}^{(0)}) \bar{\Delta}_{WB'}^{(0)} e^{-2Q b' g_{B'}^{(0)}};
\nonumber\\
\\
\Theta_{12}^{(R)} &\!\!=\!\!& \bar{\Delta}_{WB'}^{(0)}(1-B_{12}\bar{\Delta}_{WB'}^{(0)}) + (B_{12}-\bar{\Delta}_{WB'}^{(0)}) e^{-2Q b' g_{B'}^{(0)}};
\nonumber\\
\\
\Theta_{21}^{(R)} &\!\!=\!\!& B_{21} - (A_{0,22}^N + B_{22}) \bar{\Delta}_{WB'}^{(0)} 
\\
&&+ \bar{\Delta}_{WB'}^{(0)}(A_{0,22}^{N} + B_{22} - B_{21} \bar{\Delta}_{WB'}^{(0)}) e^{-2Q b' g_{B'}^{(0)}};
\nonumber
\\
\Theta_{22}^{(R)} &\!\!=\!\!& (A_{0,22}^{N} + B_{22} - B_{21} \bar{\Delta}_{WB'}^{(0)}) e^{-2Q b' g_{B'}^{(0)}} 
\nonumber\\
&&+ \bar{\Delta}_{WB'}^{(0)} (B_{21}-(A_{0,22}^N + B_{22}) \bar{\Delta}_{WB'}^{(0)})
\\
\Theta_{11}^{(L)} &\!\!=\!\!& 1 - \bar{\Delta}_{W B_1}^2 e^{-2Q b' g_{B_1}}; 
\\
\Theta_{12}^{(L)} &\!\!=\!\!& (1 - e^{-2Q b' g_{B_1}}) \bar{\Delta}_{W B_1};
\\
\Theta_{21}^{(L)} &\!\!=\!\!& -(1-e^{-2Q b' g_{B_1}}) \bar{\Delta}_{W B_1};
\\
\Theta_{22}^{(L)} &\!\!=\!\!& e^{-2Q b' g_{B_1}} - \bar{\Delta}_{W B_1}^2
\ea
\end{subequations}

\section{Two interacting multilayers with rotating optic axes} 
\label{app:multilayers}
The matrix elements of $\widetilde{\mathbf{A}}^{(j)}$ in Eq.~(\ref{eq:tAj_matrix}) are given by 
\begin{subequations}
\label{eq:thales}
\ba
\widetilde{A}_{11}^{(j)} &=& 1 - e^{-2Q a' \widetilde{g}_{B'}^{(j)}} (\widetilde{\Delta}_{WB'}^{(j)})^2,
\\
\widetilde{A}_{12}^{(j)} &=& (1 - e^{-2Q a' \widetilde{g}_{B'}^{(j)}}) \widetilde{\Delta}_{WB'}^{(j)},
\\
\widetilde{A}_{21}^{(j)} &=& - e^{-2Q a} (1 - e^{-2Q a' \widetilde{g}_{B'}^{(j)}}) \widetilde{\Delta}_{WB'}^{(j)},
\\
\widetilde{A}_{22}^{(j)} &=& e^{-2Q a}(e^{-2Q a' \widetilde{g}_{B'}^{(j)}} - (\widetilde{\Delta}_{WB'}^{(j)})^2)
\ea 
\end{subequations}
\bing{The elements of the matrix $\widetilde{\mathbf{A}}_0$ in Eq.~(\ref{eq:anne0}) are given by}
\be
\label{app:widetildeA0}
\bing{\widetilde{A}_{0,11}=1, \widetilde{A}_{0,12}=\widetilde{A}_{0,21}=0, \widetilde{A}_{0,22}=e^{-2Q(a+a')},}
\ee 
whilst those of matrices $\delta\widetilde{\mathbf{A}}^{(j)}$ (Eq.~(\ref{eq:anne0})) and $\widetilde{\mathbf{B}}$ (Eq.~(\ref{eq:caroline0})) are given by Eqs.~(\ref{eq:anne}) and (\ref{eq:caroline}): 
\begin{subequations}
\label{eq:anne}
\ba
\delta \widetilde{A}_{11}^{(j)} &\!=\!& 0,
\\
\delta \widetilde{A}_{12}^{(j)} &\!=\!& -\frac{\gamma_n}{2} (\cos(\theta_d-j\delta\theta-\psi))^2 e^{-Q a'}\sinh(Q a'),
\nonumber\\
\\
\delta \widetilde{A}_{21}^{(j)} &\!=\!& \frac{\gamma_n}{2} (\cos(\theta_d-j\delta\theta-\psi))^2 e^{-Q(2a+a')} \sinh(Q a'),
\nonumber\\
\\
\delta \widetilde{A}_{22}^{(j)} &\!=\!& -\gamma_n (\cos(\theta_d-j\delta\theta-\psi))^2 Q a' e^{-2Q(a+a')}
\ea
\end{subequations}
\begin{subequations}
\label{eq:caroline}
\ba
\widetilde{B}_{11} &=& 0, 
\\
\widetilde{B}_{12} &=&
-\frac{\gamma_n}{2} 
e^{-Q a'}
 \sinh(Q a') 
\nonumber\\ 
&&\quad \times 
V_{\bing{N_L}}(Q(a+a'), \delta\theta, \theta_d-\psi),
\\
\widetilde{B}_{21} &=& 
\frac{\gamma_n}{2} 
e^{Q a' -2\bing{N_L} Q(a+a')}
\sinh(Q a') 
\nonumber\\
&&\times   
V_{\bing{N_L}} (-Q(a+a'), \delta\theta, \theta_d-\psi),
\nonumber\\
\\
\widetilde{B}_{22} &=& 
-\frac{\gamma_n}{2} 
Q a' \, e^{-2\bing{N_L} Q(a+a')} (\bing{N_L}+\widetilde{P}_{\bing{N_L}}(\delta\theta))
\ea
\end{subequations}
In the above, the functions $V_{N}(t,\delta\theta,\psi)$ is defined by Eq.~(\ref{eq:VN}), and $\widetilde{P}_N(\delta\theta)$ is defined by
\ba
\widetilde{P}_N(\delta\theta) &\!\equiv\!&
2\sum_{j=1}^{N} (\cos(\theta_d-j\delta\theta-\psi))^2 - N
\nonumber\\
&\!=\!& 
\cos(2\theta_d\!-\!(N+1)\delta\theta \!-\! 2\psi) \frac{\sin N\delta\theta}{\sin \delta\theta}
\ea
The matrix elements of $\mathbf{\Theta}^{(R)}$ and $\mathbf{\Theta}^{(L)}$ in Eqs.~(\ref{eq:multilayerM}) are given by
\begin{subequations}
\label{eq:multilayerelms}
\ba
\Theta_{11}^{(R)} &=& 1 - B_{12} \bar{\Delta}_{WB'}^{(0)} + (B_{12} - \bar{\Delta}_{WB'}^{(0)}) \bar{\Delta}_{WB'}^{(0)} e^{-2Q b' g_{B'}^{(0)}};
\nonumber\\
\\
\Theta_{12}^{(R)} &=& \bar{\Delta}_{WB'}^{(0)}(1-B_{12}\bar{\Delta}_{WB'}^{(0)}) + (B_{12}-\bar{\Delta}_{WB'}^{(0)}) e^{-2Q b' g_{B'}^{(0)}};
\nonumber\\
\\
\Theta_{21}^{(R)} &=& B_{21} - (\bing{A_{0,22}^{N_R}} + B_{22}) \bar{\Delta}_{WB'}^{(0)} 
\nonumber\\
&&+ \bar{\Delta}_{WB'}^{(0)}(\bing{A_{0,22}^{N_R}} + B_{22} - B_{21} \bar{\Delta}_{WB'}^{(0)}) e^{-2Q b' g_{B'}^{(0)}};
\\
\Theta_{22}^{(R)} &=& (\bing{A_{0,22}^{N_R}} + B_{22} - B_{21} \bar{\Delta}_{WB'}^{(0)}) e^{-2Q b' g_{B'}^{(0)}} 
\nonumber\\
&&+ \bar{\Delta}_{WB'}^{(0)} (B_{21}-(\bing{A_{0,22}^{N_R}} + B_{22}) \bar{\Delta}_{WB'}^{(0)})
\\
\Theta_{11}^{(L)} &=& 1 + \widetilde{B}_{21} \widetilde{\Delta}_{WB'}^{(0)} 
- \widetilde{\Delta}_{WB'}^{(0)} (\widetilde{B}_{21}
+\widetilde{\Delta}_{WB'}^{(0)}) e^{-2Q \bing{a'} \widetilde{g}_{B'}^{(0)}}; 
\nonumber\\
\\
\Theta_{12}^{(L)} &=& \widetilde{B}_{12} + \widetilde{\Delta}_{WB'}^{(0)} 
(\bing{\widetilde{A}_{0,22}^{N_L}} + \widetilde{B}_{22}) 
\nonumber\\
&&- \widetilde{\Delta}_{WB'}^{(0)} 
(\widetilde{B}_{12} \widetilde{\Delta}_{WB'}^{(0)} + \bing{\widetilde{A}_{0,22}^{N_L}} + \widetilde{B}_{22})
e^{-2Q \bing{a'} \widetilde{g}_{B'}^{(0)}}
\\
\Theta_{21}^{(L)} &=& - \widetilde{\Delta}_{WB'}^{(0)} (1 + \widetilde{B}_{21} \widetilde{\Delta}_{WB'}^{(0)}) + (\widetilde{B}_{21} + \widetilde{\Delta}_{WB'}^{(0)}) e^{-2Q \bing{a'} \widetilde{g}_{B'}^{(0)}}
\nonumber\\
\\
\Theta_{22}^{(L)} &=&  (\bing{\widetilde{A}_{0,22}^{N_L}} + \widetilde{B}_{22} + \widetilde{B}_{12} \widetilde{\Delta}_{WB'}^{(0)}) 
e^{-2Q \bing{a'} \widetilde{g}_{B'}^{(0)}}
\nonumber\\
&&- \widetilde{\Delta}_{WB'}^{(0)} 
(\widetilde{B}_{12} + (\bing{\widetilde{A}_{0,22}^{N_L}} + \widetilde{B}_{22}) \widetilde{\Delta}_{WB'}^{(0)})
\ea
\end{subequations}
\bing{In the above, the values of $B_{12}$, $B_{21}$ and $B_{22}$ are given by Eqs.~(\ref{eq:Bele}) with $N \rightarrow N_R$.} 
In the weak anisotropy regime, the effective reflection coefficients $\bar{\Delta}_{WR}^{{\rm (eff)}}$ and $\bar{\Delta}_{W B_1}^{{\rm (eff)}}$ defined by Eq.~(\ref{eq:yael}) can be approximated by 
\begin{subequations}
\label{eq:victoria}
\ba
\bar{\Delta}_{WR}^{{\rm (eff)}} 
&\approx& 
-\frac{\gamma_n}{2}e^{-Q b'}\sinh(Q b')\cos^2\psi 
\nonumber\\
&&- \frac{\gamma_n e^{-Q b'} \sinh(Q b')}{8(\cosh(2Q(b+b'))-\cos(2\delta\theta))}
\nonumber\\
&&\quad\times 
\big[ 
\cos(2(\delta\theta-\psi)) 
- e^{-2Q(b+b')}\cos(2\psi)
\nonumber\\
&&\qquad+ 
e^{-2(N_R+1)Q(b+b')} \cos(2(N_R\delta\theta-\psi))
\nonumber\\
&&\qquad-
e^{-2N_R Q(b+b')} \cos(2((N_R+1)\delta\theta-\psi))
\big]
\nonumber\\
&&-\frac{\gamma_n}{4} e^{-Q((N_R+1)b+(N_R+2)b')} \sinh(Q b') 
\nonumber\\
&&\quad\times
\frac{\sinh(N_R Q(b+b'))}{\sinh(Q(b+b'))}
\ea
\ba
\bar{\Delta}_{W B_1}^{{\rm (eff)}} 
&\approx& 
-\frac{\gamma_n}{2} e^{-Q a'} \sinh(Q a') \cos^2(\theta_d-\psi)
\nonumber\\
&&-
\frac{\gamma_n e^{-Q a'} \sinh(Q a')}{8(\cosh(2Q(a+a'))-\cos(2\delta\theta))}
\nonumber\\
&&\quad\times 
\big[ 
\cos(2(\delta\theta-\theta_d+\psi)) 
\nonumber\\
&&\qquad- 
e^{-2Q(a+a')} \cos(2(\theta_d-\psi))
\nonumber\\
&&\qquad+
e^{-2(N_L+1)Q(a+a')} \cos(2(N_L\delta\theta - \theta_d + \psi))
\nonumber\\
&&\qquad- 
e^{-2Q N_L(a+a')} \cos(2((N_L+1)\delta\theta-\theta_d+\psi))
\big]
\nonumber\\
&&-
\frac{\gamma_n}{4} e^{-Q((N_L+1)a+(N_L+2)a')} \sinh(Q a') 
\nonumber\\
&&\quad\times
\frac{\sinh(N_L Q(a+a'))}{\sinh(Q(a+a'))}
\ea
\end{subequations}
The coefficients $h_0$, $h_1$, $h_2$ and $h_3$ of $G_{rr}$ in Eq.~(\ref{eq:Gmulti}) are given by
\begin{widetext}
\begin{subequations}
\label{eq:endrina}
\ba
h_0 &\equiv& \frac{\pi e^{Q(a+b)} \sinh(Q a')\sinh(Q b')}{32\sinh(Q(a+a'))\sinh(Q(b+b'))},
\\
h_1 &\equiv& \frac{\pi e^{-Q(a'+b')} \sinh(Q a')\sinh(Q b') (e^{2Q(a+a')} - 2\cos(2\delta\theta)) (e^{2Q(b+b')} - 2\cos(2\delta\theta))}{64 (\cos(2\delta\theta) - \cosh(2Q(a+a'))) (\cos(2\delta\theta) - \cosh(2Q(b+b')))},
\\
h_2 &\equiv& \frac{\pi e^{-Q(a'+b')} (e^{2Q(a+a')} + e^{2Q(b+b')} - 4 \cos(2\delta\theta)) \sinh(Q a')\sinh(Q b')}{64 (\cos(2\delta\theta) - \cosh(2Q(a+a'))) (\cos(2\delta\theta) - \cosh(2Q(b+b')))},
\\
h_3 &\equiv& \frac{\pi e^{-Q(a'+b')} \sinh(Q a')\sinh(Q b') }{64 (\cos(2\delta\theta) - \cosh(2Q(a+a'))) (\cos(2\delta\theta) - \cosh(2Q(b+b')))}.
\ea
\end{subequations}
\end{widetext}
For the case where the optic axes within each slab are all aligned (i.e., $\delta\theta=0$), we obtain 
\be
h_1+h_2+h_3 = \frac{\pi e^{Q(a+b)} \sinh(Q a') \sinh(Q b')}{64 \sinh(Q(a+a')) \sinh(Q(b+b'))}
\ee
which combined with Eq.~(\ref{eq:Gmulti}), yields Eq.~(\ref{eq:Grr0}).


\begin{thebibliography}{99}
\bibitem{footnote:matching}
\bing{In principle one needs to take all Matsubara frequencies into account. The dielectric response of the uniaxial material is also different from that of the solvent. However, if the static values for the two materials are approximately the same, and the dielectric function undergoes a monotonic decay to zero with increasing frequency for imaginary frequencies, the difference between the dielectric response of the two materials should not be significant. The difference becomes even smaller for higher temperatures, where the static, zero frequency contribution dominates over the other contributions. For systems where there is no dielectric matching between the ordinary axes of the uniaxial layers and the solvent medium, our approach can be generalized by attaching suitable prefactors to $g_{B_1}$ and $g_{B_2}$ in the reflection coefficients of Eqs.~(\ref{eq:Deltas}): $g_{B_1} \rightarrow (\varepsilon_{B_1,z}/\varepsilon_{W}) g_{B_1}, \, g_{B_2} \rightarrow (\varepsilon_{B_2,z}/\varepsilon_{W}) g_{B_2}.$}
\end{thebibliography}
\end{document}